\documentstyle[tighten,floats,aps,epsf]{revtex}

\begin{document}
\title
{Cluster Structures of the Ground and Excited States of $^{12}$Be 
Studied with Antisymmetrized Molecular Dynamics}

\author{Y. Kanada-En'yo}

\address{Institute of Particle and Nuclear Studies, \\
High Energy Accelerator Research Organization,\\
Ibaraki 305-0801, Japan}

\author{H. Horiuchi}

\address{Department of Physics, Kyoto University, Kyoto 606-8502, 
Japan}

\maketitle
\begin{abstract}
The structures of the ground and excited states of $^{12}$Be 
were studied with antisymmetrized molecular dynamics.
The ground state was found to be a state with 
a developed 2$\alpha$ core with two neutrons occupying the intruder orbits. 
The energy levels of the newly measured spin-assigned states 
were described well, except for the $1^-_1$ state.
The calculations indicated that many exotic cluster structures appear
in the low-energy region. The widths concerning
 $\alpha$ and $^6$He decays were discussed by using reduced width 
amplitudes.
\end{abstract}

\noindent

\section{Introduction}
Owing to the progress of experimental techniques, 
information concerning the excited states of light unstable nuclei 
has rapidly increased.
Recently, exotic clustering in the 
light unstable nuclei has become one of the attractive subjects in
3experimental and theoretical research.
Since, in light stable nuclei, it has already been known that 
clustering is one of the essential features of nuclear dynamics, 
not only in excited
states, but also in ground states, it is natural to expect
cluster features in light unstable nuclei.
Pioneering theoretical studies have suggested
the development of cluster structures
with a 2$\alpha$ core in Be and B isotopes 
\cite{SEYA,ENYObc,OERTZEN,ARAI,DOTE,ENYOf,ENYOg,ITAGAKI,ITAGAKIa,OGAWA,ENYObe11,ENYObe14}.
Especially, highly excited states with developed cluster structures 
in $^{10}$Be
and $^{11}$Be have been studied by microscopic calculations 
\cite{ENYOf,ENYOg,ITAGAKI,ITAGAKIa,OGAWA,ENYObe11}.

In the case of $^{12}$Be, 
the existence of cluster states was suggested in experimental
measurements of the excited states \cite{TANIHATA}. 
In recent experiments of $^6$He+$^6$He and $^8$He+$^4$He 
breakup reactions \cite{FREER,SAITO}, 
many new excited states were discovered above the threshold energies.
Some of the states are candidates of exotic cluster states,
since the measured spin-parities of those excited states indicate 
a rotational band with a large moment of inertia.
It is an interesting subject
to investigate clustering aspects in $^{12}$Be.
Although the molecular states in $^{12}$Be
were theoretically suggested with a potential
model with He clusters \cite{ITO}, they have not yet been studied 
by microscopic calculations. In the recent investigations
with the Generator Coordinate Method of a coupled-channel
 two-body cluster model, 
using microscopic $^6$He+$^6$He and $^8$He+$^4$He wave functions,
a strong mixing of both configurations was predicted 
\cite{DESCOUVEMENT}.

The ground and the low-lying states of $^{12}$Be present us with 
other attractive
subjects concerning the vanishing of a neutron magic number, 8.
The vanishing of the neutron magic number is already known in a 
neighboring nucleus, $^{11}$Be.
The vanishing in $^{12}$Be was predicted in the
early theoretical works \cite{BARKER}.
The abnormal configurations($(sd)^2$) in the low-lying states
were experimentally studied by $^{10}$Be$(t,p)^{12}$Be reaction
\cite{FORTUNE}.
The abnormal structure of the ground state
was also suggested in the analyses of the beta-decay strength 
from the ground state of $^{12}$Be
into $^{12}$B \cite{BARKER,FORTUNE,TSUZUKIa}. 
Moreover, the vanishing in $^{12}$Be was 
supported by a recent measurement of the spin-parity for 
a low-lying $1^-$ state \cite{IWASAKI}.
A new low-lying $0^+_2$ state \cite{SHIMOURA} is also 
important to solve the inversion mechanism.
In the recent theoretical studies, 
the structure of low-lying states were 
microscopically described by a molecular-orbital model
by N. Itagaki et al. \cite{ITAGAKIa}, and by a method of antisymmetrized 
molecular dynamics by one of the authors (Y.K.) \cite{ENYObe14}.
They were also discussed by F. M. Nunes et al. within a three-body 
(n+n+$^{10}$Be) model with
core excitation \cite{NUNES}. 

Our aim is to conduct systematic research of the structures of the
ground and excited states of $^{12}$Be based on microscopic theoretical 
calculations while focusing on clustering aspects.
First of all, we have studied the systematics of the level 
structures, including the 
experimentally observed excited states. We have searched
 for cluster and non-cluster
states to solve the following problems concerning clustering
aspects in $^{12}$Be.
Do their cluster structures appear in $^{12}$Be ? 
If they do appear, what are the characteristics 
of their structures in unstable nuclei compared
with those of stable nuclei ?
The roles of the valence nucleons in the cluster states are 
interesting problems. We have also investigated the mechanism of 
the development and breaking of clustering.

The important point is that the theoretical approach should be free from 
model assumptions, such as the stability of the mean-field 
and the existence of inert cores or clusters, because we have to describe 
various structures covering developed cluster structures
as well as shell-model-like structures in 
the ground and excited states.
It is difficult to study developed clustering in 
excited states with such mean-field approaches as the traditional 
shell model and the Hartree-Fock model. 
The cluster structures of the excited states of 
$^9$Be and $^{10}$Be were successfully explained by cluster models 
\cite{SEYA,ARAI,ITAGAKI,OGAWA,OKABE} while
assuming a 2$\alpha$ core and surrounding neutrons.
However, we think that a description based on 
cluster models simplifying the system as a 2$\alpha$ core
with valence neutrons or 2-He 
clusters is not sufficient for a systematic 
investigation of $^{12}$Be because of many valence neutrons in the system.
We have applied a theoretical approach of 
antisymmetrized molecular dynamics (AMD). 
The AMD method has already proved to be a useful theoretical approach
for the nuclear structure \cite{ENYObc,DOTE,ENYOg,ENYObe11,ENYOa,ENYOe}.
Within the AMD framework, we do not need such model assumptions as
inert cores, clusters, or axial symmetries, because the 
wave function of the nuclear system is written by Slater determinants,
where the spatial part of each single-particle wave function 
for a nucleon is expressed by a localized Gaussian wave packet. 
Due to the flexibility 
of the AMD wave function, we successfully described the structure changes 
between shell-model-like states and cluster states as functions of 
the neutron number in light unstable nuclei. 
Owing to progress in computational power, it has become possible to study
excited states by extended AMD calculations. This method is
based on variational calculation after spin-parity projection
(VAP calculation) within the framework of AMD, which has already been
confirmed to be powerful for studying the excited states of light nuclei,
as shown in studies of the stable nucleus $^{12}$C \cite{ENYOe}, and 
also unstable nuclei($^{10}$Be and $^{11}$Be) \cite{ENYOg,ENYObe11}.
The authors and their collaborators succeeded to 
describe various structures of the excited states and 
to reproduce many kinds of experimental data for 
nuclear structures of these nuclei with the AMD method.

In the present work, the structures of the ground and excited states 
of $^{12}$Be
were analyzed by performing a variational calculation after 
spin-parity projection based on the AMD method.
In the next section (Sec. \ref{sec:formulation}), 
we explain the formulation of AMD for a study of the nuclear
structure of excited states. The adopted effective 
interactions are explained in Sec. \ref{sec:interaction}.  
In Sec. \ref{sec:results}, we present the calculated results concerning such
observables as the energy levels, radii and $\beta$ decays as well as the
 $E1$ and $E2$ transitions compared with the experimental data.
In the discussion(Sec. \ref{sec:discuss}), the intrinsic structures 
and the rotational band structures are described.
The single-particle behavior of the valence neutrons is analyzed.
We discuss the systematics of the development of cluster states 
in Be isotopes, which can be classified according to the neutron 
orbits surrounding the 2$\alpha$ core.
We investigate the inter-cluster motions between He clusters extracted from 
the obtained $^{12}$Be wave functions, and 
calculate the decay widths for the $^{4}$He and $^{6}$He channels
by using the method of reduced width amplitudes.

\section{Formulation}
 \label{sec:formulation}

The formulation of AMD 
for a nuclear structure study of ground and excited states
is explained in \cite{ENYObc,ENYOg,ENYOe}.
In particular, the formulation of the present calculations
is same as that described in Ref. \cite{ENYOg}.
In this section, I briefly review the formulation.

The wave function of a system is written by a superposition of 
the AMD wave functions($\Phi_{AMD}$). 
The AMD wave function of a nucleus with a mass number $A$
is a Slater determinant of Gaussian wave packets.
The $i$th single-particle wave function 
is a product of the spatial wave function,
 the intrinsic spin function  and 
the iso-spin function.
The spatial part is presented by 
variational complex parameters, ${\rm X}_{1i}$, ${\rm X}_{2i}$, 
${\rm X}_{3i}$, which indicate the center of the Gauusian wave packets.
The orientation of the intrinsic spin part is expressed by
a variational complex parameter $\xi_{i}$, and the iso-spin
function is fixed to be up(proton) or down(neutron)
in the present calculations.
Thus, an AMD wave function is expressed by a set of variational parameters,
${\bf Z}\equiv \{{\rm X}_{ni},\xi_i\}\ (n=1,2,3\ \hbox{and }  i=1,\cdots,A)$, 
which indicate the centers of Gaussians of the spatial part 
and the spin orientations of the intrinsic spin part of the single-particle
wave functions.

When we consider a parity-eigen state projected from an AMD wave function,
the total wave function is written by two Slater determinants.
In the case of a total-angular-momentum eigen state,
the wave function of the state is represented by the integral of 
the rotated AMD wave functions.
The expectation values of a given tensor operator for the 
total-angular-momentum projected states 
are calculated 
by evaluating the integral with a sum 
over mesh points of the Euler angles.

In principle, the total wave function can be a superposition of independent
AMD wave functions. 
We can make the superposition 
of the spin-parity projected AMD wave functions
($P^{J\pm}_{MK'}\Phi_{AMD}$)
as follows:
\begin{equation}
\Phi=cP^{J\pm}_{MK'}\Phi_{AMD}({\bf Z})
+c'P^{J\pm}_{MK'}\Phi_{AMD}({\bf Z}')+\cdots.
\end{equation}

We performed a variational calculation for a trial wave function
to find the state which minimizes the energy of the system,
\begin{equation}
\frac{\langle\Phi|H|\Phi\rangle}{\langle\Phi|\Phi\rangle}
\end{equation}
by the method of frictional cooling, which is one of the imaginary 
time methods.
Regarding the frictional cooling method in AMD, the reader is referred to 
two papers \cite{ENYObc,ENYOa}.

In order to obtain the wave function for the lowest $J^\pm$ state,
we varied the parameters ${\bf X}_i$ and $\xi_{i}$($i=1\sim A$) to
minimize the energy expectation value of a
spin-parity projected AMD wave function,
$\Phi=P^{J\pm}_{MK'}\Phi_{AMD}({\bf Z})$,
by using the frictional cooling method.
That is to say, we performed energy variation after 
spin-parity projection(VAP) for an AMD wave function. 

With the VAP calculation for the $J^\pm$ eigen state,
$\Phi^{J\pm}_1({\bf Z})=P^{J\pm}_{MK'}\Phi_{AMD}({\bf Z})$,
 with an appropriate $K'$,
we obtained a set of parameters, ${\bf Z}={\bf Z}^{J\pm}_1$, which presents
the wave function of the first $J^\pm$ state. 
In order to search for the parameters ${\bf Z}={\bf Z}^{J\pm}_n$ of the 
$n$th $J^\pm$ state,
the wave functions were superposed so as to be orthogonal
to the lower states. The parameters ${\bf Z}^{J\pm}_n$ 
for the $n$th $J^\pm$ state were provided by varying ${\bf Z}$ 
so as to minimize the energy of the wave function orthogonalized 
to the lower states.

After the VAP calculation of the $J^\pi_n$ states for various
$J$, $n$ and $\pi=\pm$,
we obtained the optimum  intrinsic states,
$\Phi_{AMD}({\bf Z}^{J\pi}_n)$, 
which approximately describe the corresponding $J^\pi_n$ states. 
In order to obtain more precise wave functions, we superposed the 
spin-parity eigen wave functions projected from all of the obtained 
intrinsic states.
Namely, we determined the final wave functions for the $J^\pm_n$ states
by simultaneously diagonalizing the Hamiltonian matrix, 
$\langle P^{J\pm}_{MK'} \Phi_{AMD}({\bf Z}^{J_i \pi_i}_{n_i})
|H|P^{J\pm}_{MK''} \Phi_{AMD}({\bf Z}^{J_j \pi_j}_{n_j})\rangle$,
and the norm matrix,
$\langle P^{J\pm}_{MK'} \Phi_{AMD}({\bf Z}^{J_i\pi_i}_{n_i})
|P^{J\pm}_{MK''} \Phi_{AMD}({\bf Z}^{J_j\pi_j}_{n_j})\rangle$,
with regard to ($i,j$) for 
all of the obtained intrinsic states, and 
to ($K', K''$).
Compared with the experimental data, such as the energy levels and
 $E2$ transitions, the expectation values were calculated with the 
final states after diagonalization.

\section{Interactions} 
\label{sec:interaction}

The adopted interaction is the sum of the central force, the
 spin-orbit force and the Coulomb force.
The central force is chosen to be the 
MV1 force of case 3 \cite{TOHSAKI},
which contains a zero-range three-body force, $V^{(3)}$, 
as a density-dependent 
term in addition to the two-body interaction, $V^{(2)}$, of 
the modified Volkov No.1 force.
The spin-orbit force of the G3RS force \cite{LS} is adopted.

\section{Results}\label{sec:results}

The structures of the excited states of $^{12}$Be were studied 
based on the VAP calculation within the framework of AMD.
In this section, we present theoretical results, such as the 
energy levels, radii, and transitions ($\beta$, $E2$, $E1$ and $E0$), 
while comparing them with the experimental data.
A detailed analysis of the structures of the states
is given in the next section.
 
The adopted interaction parameters in the present work are those used 
in Refs.\cite{ENYOg,ENYObe11}, which reproduced 
the abnormal spin-parity 1/2$^+$ of the ground state of $^{11}$Be.
Namely, the Majorana, Bartlett and Heisenberg 
parameters in the central force are $m=0.65$, $b=h=0$, 
and the strength of the spin-orbit force is chosen to be 
$u_I=-u_{II}=3700$ MeV.
The width parameter ($\nu$) was chosen to be 0.17 fm$^{-2}$, which gave a
minimum energy of $^{12}$Be in a simple AMD calculation without the 
spin projection.

The wave functions for the lowest $J^\pm$ states were obtained by 
the VAP calculation of $P^{J\pm}_{MK'}\Phi_{AMD}$ by choosing $(J^\pm,K')$=
$(0^+,0)$, $(2^+,0)$, $(4^+,0)$, $(6^+,0)$, $(8^+,0)$,
$(1^-,1)$, $(2^-,1)$, $(3^-,1)$, $(4^-,1)$, $(5^-,1)$
$(6^-,1)$, $(3^+,2)$, $(5^+,2)$, $(7^+,2)$, $(0^-,0)$, $(1^+,1)$,
 $(1^+,0)$. 
After obtaining the lowest states($J^\pm_1$), we calculated the 
second and third $J^\pm$ states ($0^+_2$, $0^+_3$, 
$2^+_2$, $2^+_3$, $4^+_2$, $6^+_2$, $1^-_2$)
with the VAP calculation of the superposed wave functions
orthogonal to the obtained lower $J^\pm$ states.
The obtained AMD wave functions are considered to approximately 
describe the intrinsic states of the corresponding $J^\pm_n$ states.
The final wave functions of the $J^\pm_n$ states were determined by 
superposing the spin-parity eigen states projected from these obtained
AMD wave functions so as to simultaneously diagonalize the Hamiltonian
matrix and the norm matrix. 
In principle, the number of the superposed AMD wave functions is 24, 
which is the number of calculated levels. In the present 
calculations, we omitted several of them which are not important,
just to save computational time.
Namely, we diagonalize the positive-parity(negative-parity) states 
projected from 22(18) AMD wave functions. 

\subsection{Energies}\label{subsec:energy}

The theoretical binding energy of $^{12}$Be was found to be 61.9 MeV(60.4 MeV) 
after(before) diagonalization, which underestimated the experimental
value, 68.65 MeV.
We found that the binding could be improved by changing the Majorana
parameter ($m$) of the interaction to a smaller value.
For example, a theoretical value of 66.1 MeV was obtained by using $m=0.62$
with a simple VAP calculation before diagonalization. 
Nevertheless, we adopted the parameter $m=0.65$ in the present work, 
because this value reproduces the parity inversion of $^{11}$Be
\cite{ENYObe11}, which 
should be important for describing the ground-state properties of 
the neighboring nucleus, $^{12}$Be.
We checked that the change in the $m$ parameter has no 
significant effect on the excitation energies, $E(0^+_2)$ and $E(8^+_1)$,
 of $^{12}$Be, at least. In fact, the calculated excitation 
energies of $E(0^+_2)$ before
diagonalization are 3.0 MeV and 2.8 MeV, 
in the cases of $m=0.62$ and $m=0.65$, respectively.

\begin{figure}
\noindent
\epsfxsize=0.49\textwidth
\centerline{\epsffile{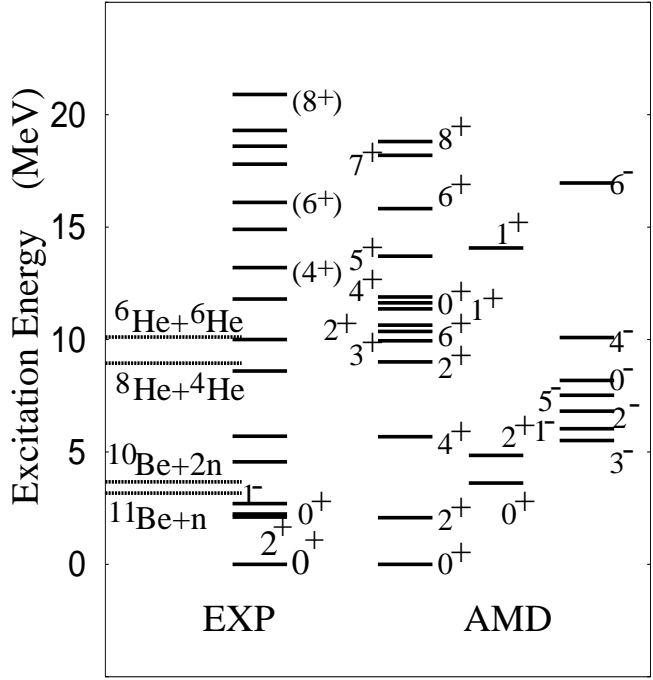}}
\caption{\label{fig:be12sped}
Excitation energies of the levels in $^{12}$Be. 
The experimental data are from the Table of Isotopes and Refs. 
\protect{\cite{FREER,IWASAKI,TANIHATA,SAITO,SHIMOURA}}.}
\end{figure}

The theoretical level scheme 
is shown in Fig.\ref{fig:be12sped}. 
The calculations suggest that many excited states appear in the 
low-energy region. 
This is the first theoretical work which systematically
reproduces the experimental energy levels of all the spin-assigned states,
 except for the $1^-$ state.
By analyzing the intrinsic AMD wave functions of the states, 
we can consider that there exist 
rotational bands ($K^\pi$=$0^+_1$, $0^+_2$, $0^+_3$
and $1^-_1$), which consist of the following states: 
\{$0^+_1$, $2^+_1$, $4^+_1$, $6^+_1$ and $8^+_1$\},
\{$0^+_2$, $2^+_2$\},
\{$0^+_3$, $2^+_4$, $4^+_2$ and $6^+_2$\}
and \{$1^-_1$, $2^-_1$, $3^-_1$, $4^-_1$, $5^-_1$\}, respectively

\begin{figure}
\noindent
\epsfxsize=0.49\textwidth
\centerline{\epsffile{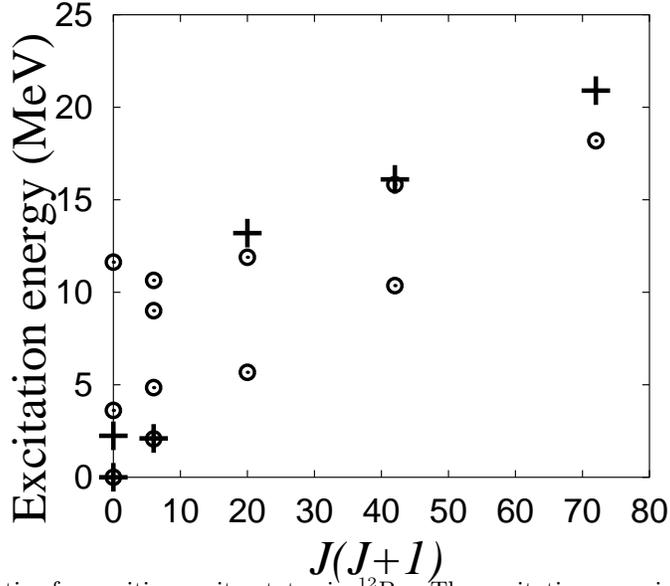}}
\caption{\label{fig:be12rot}
Energy-spin systematics for positive-parity states in $^{12}$Be.
The excitation energies of the natural spin states are plotted as 
functions of $J(J+1)$. The circle symbols are the 
theoretical results, while the cross symbols are 
the experimental data of the spin-parity assigned states, 
which are taken from the Table of Isotopes and 
Refs. \protect{\cite{FREER,SHIMOURA}}.}
\end{figure}

The energy-spin systematics for positive-parity states with natural spins
are shown in Fig. \ref{fig:be12rot}.
It is surprising that the newly observed levels, 
$4^+$ at 13.2 MeV and $6^+$ at 16.1 MeV \cite{FREER}, correspond  well
to the $4^+_2$ and $6^+_2$ states obtained in the present results.
An interesting point is that these excited states
belong not to the yrast band, $K^\pi=0^+_1$, but to the new excited
$K^\pi=0^+_3$ band with developed cluster structure.
The theoretical results predict 
the existence of many positive-parity states belonging to 
the $K^\pi=0^+_1$ and $K^\pi=0^+_2$ bands in the lower energy region.
Even though $^{12}$Be has a neutron magic number of 8, 
the calculated ground $K^\pi=0^+$ band has a large moment of inertia,
because the band head $0^+_1$ state is 
not the ordinary state with a closed neutron $p$-shell, but an intruder 
state with a developed cluster structure. 
It has a prolate deformed structure, which is dominated by 
$2p$-$2h$ configurations, and reaches the band terminal at the $8^+_1$ state,
because the $J^\pm=8^+_1$ state is the highest-spin state in the 2$\hbar\omega$
configurations.
The third $0^+$ band ($K^\pi=0^+_3$)
was found to be represented by other $2p$-$2h$ configurations than those in 
the $K^\pi=0^+_1$ band. The $K^\pi=0^+_3$ band 
has an extremely large moment of inertia because of 
a remarkably developed $^6$He+$^6$He clustering, and reaches the band terminal
$6^+_2$ state accompanying the spin-alignment of nucleons 
in the high-spin region.
On the other hand, the main components of the $0^+_2$ and $2^+_2$ states are 
of the $0\hbar\omega$ configurations, with the closed neutron $p$-shell,
and constitute the $K^\pi=0^+_2$ band. The band head $0^+_2$ state of 
this band, which was theoretically predicted to appear just above the
intruder ground band by Itagaki et al.\cite{ITAGAKI}
and Kanada-En'yo et al. \cite{ENYOsup}, was recently discovered in an
observation of coincidence gamma rays by Shimoura et al. \cite{SHIMOURA}.

The theoretical excitation energy of the $1^-_1$ state is larger than the
measured $1^-$ state at 2.68 MeV \cite{IWASAKI}. In spite of the 
overestimation of the excitation energy, we regard this $1^-_1$ state 
in the $K^\pi=1^-$ as the $1^-$ state at 2.68 MeV
because of the large $E1$ transition strength, as shown later.
In order to improve the excitation energy of this $1^-$
state, it is important to take the mixing of the other configuration 
into account. For example, by mixing a $1^-$ state with $K^\pi=0^-$, 
which is obtained at slightly higher energy than the $1^-$ state with
$K^\pi=1^-$ in VAP calculations, the $1^-_1$ state gains about 1 MeV.
Another reason for the overestimation may be connected to the 
fact that the present interactions give a too small value for the 
effective energy difference between 
the $s_{1/2}$ and $d_{5/2}$ orbits, which was found
in the calculations of $^{11}$Be in Ref. \cite{ENYObe11}.

\subsection{Radii}
The theoretical values of the root-mean-square radii of the density
distributions of point-like nucleons,
protons, and neutrons are listed in Table \ref{tab:radii} together with the 
experimental matter radius \cite{TAN88} deduced from the reaction 
cross sections.
Because of deformations,
the proton and neutron radii in the ground state are larger 
compared with those in the $0^+_2$ state.
Since the theoretical values concerning the difference 
$\Delta r=r_n-r_p$ between the proton and neutron radii 
are about 0.3 fm, $^{12}$Be is a candidate of the neutron skin nucleus. 
Compared with the experimental data,
the present result for $0^+_1$ is larger
because the present parameter $m=0.65$ is considered to be too large to
quantitatively reproduce the radii of $p$-shell nuclei.

\begin{figure}
\noindent
\epsfxsize=0.45\textwidth
\centerline{\epsffile{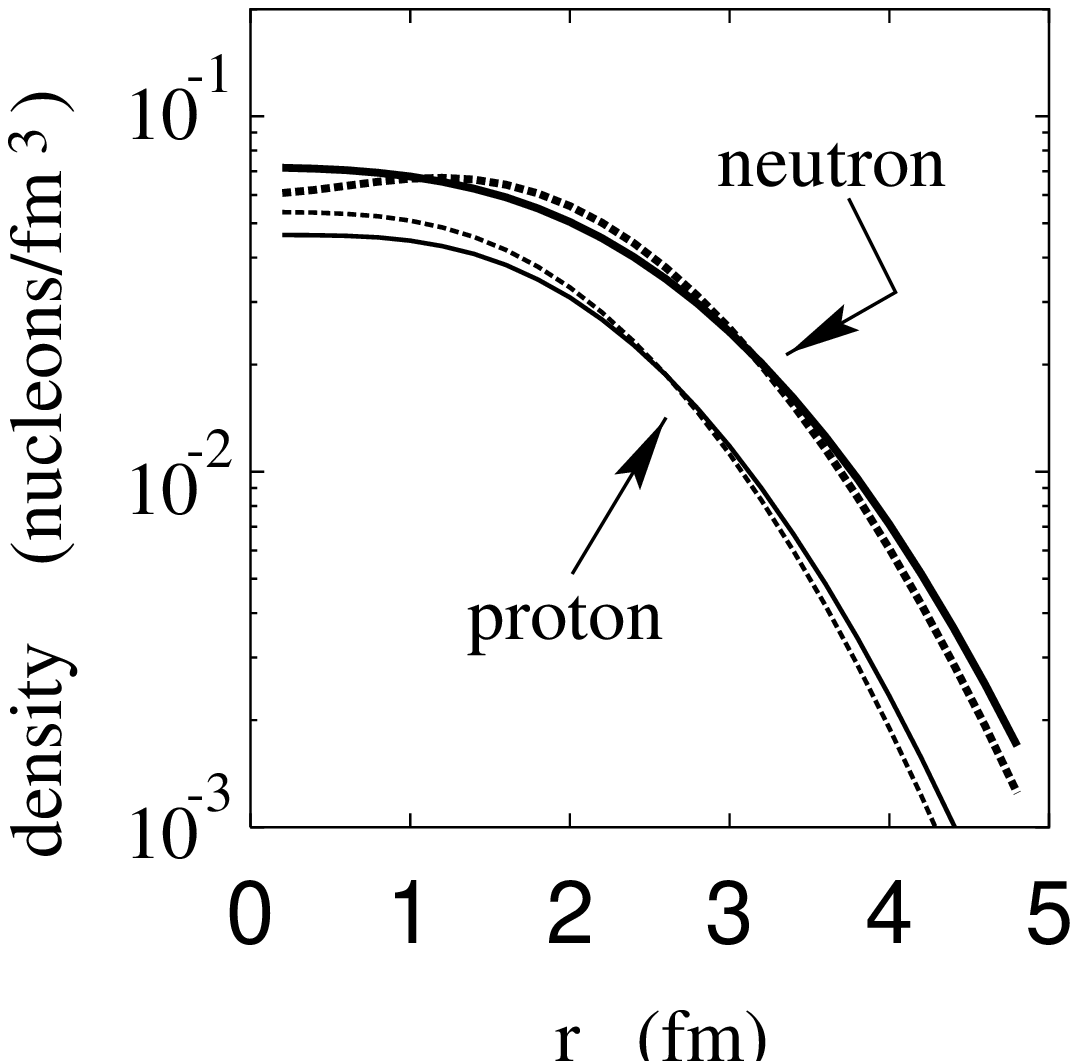}}
\caption{\label{fig:rdens}
Densities as functions of the radial coordinate ($r$).
The densities of the protons and neutrons in the $0^+_1$ state
(solid) and in the $0^+_2$ state(dashed) are shown.}
\end{figure}

The proton and neutron densities as functions of the radius 
are shown in Fig. \ref{fig:rdens}.
In both the $0^+_1$ and $0^+_2$ states, the excess neutrons enhance 
the neutron density in the surface region. In the $0^+_1$ state, 
due to cluster 
development, the proton density increases in the surface region 
while it decreases at the center of the nucleus.
It should be noted that, even if a nucleus has a neutron halo 
(the long tail of the neutron density in the outer region), 
the details of the halo structure are not expressed in the present 
model space because of a limitation of the Gaussian forms in 
AMD wave functions.

\begin{table}
\caption{ \label{tab:radii} The root-mean-square radii of 
the density distribution of point-like nucleons,
 protons, and neutrons. The experimental matter radius deduced from the 
reaction cross sections is taken from Ref.\protect\cite{TAN88}.}
\begin{center}
\begin{tabular}{cccc}
  & matter &proton&neutron\\ 
 $^{12}$Be($0^+_1$) & 2.85 fm & 2.67 fm & 2.94 fm \\
$^{12}$Be($0^+_2$) & 2.75 fm & 2.56 fm & 2.84 fm \\
\hline
exp.& $2.59\pm 0.06$ &$-$ &$-$ \\
\end{tabular}
\end{center}
\end{table}

\subsection{$\beta$ decay strength}

The strength of the $\beta$ decay 
from $^{12}$Be(0$^+_1$) to $^{12}$B($1^+$) provides helpful 
experimental evidence for 
breaking of the neutron $p$-shell closure in $^{12}$Be($0^+_1$).
T. Suzuki et al. suggested that
weak $\beta$ decay  is experimental evidence for an admixture of 
non-$0\hbar\omega$ configurations in $^{12}$Be($0^+_1$),
because the $\beta$ decays from a normal $p$-closed state of 
$^{12}$Be($0^+_1$) must be stronger.
We calculated
the Gamow-Teller transition strength,
$B(GT)\equiv |\langle \sigma\tau \rangle|^2$, where
the wave function for the daughter state, $^{12}$B($1^+_1$), was
obtained by a VAP calculation with $(J^\pm,K')=(1^+,-1)$.
The $B(GT)$ values are given shown in Table \ref{tab:be12beta}.
The present result, $B(GT)=0.9$, is almost as small as
the experimental data, $B(GT)=0.59$,
because the component of the $2\hbar\omega$ configurations
in the parent $^{12}$Be(0$^+_1$) makes the transition matrix element
of the Gamow-Teller operator to be small. 
In fact, the original $0^+_1$ state of $^{12}$Be 
obtained by a simple VAP calculation before diagonalization 
has very weak $\beta$ transitions as 
$B(GT:^{12}$Be($0^+_1)\rightarrow ^{12}$B($1^+_1))=0.2$.
On the other hand, the decay from $0^+_2$ is strong as 
$B(GT:^{12}$Be($0^+_2)\rightarrow ^{12}$B($1^+_1))=3.0$ 
before diagonalization, because the parent state
has the ordinary $0\hbar\omega$ configuration.
After state mixing with diagonalization, 
the $GT$ strength from $^{12}$Be($0^+_2$) is distributed to
that from $^{12}$Be($0^+_1$).
As a result, after the diagonalization 
the $B(GT:^{12}$Be($0^+_1)\rightarrow ^{12}$B($1^+_1))$ 
value increases to 0.9, while 
$B(GT:^{12}$Be($0^+_2)\rightarrow ^{12}$B($1^+_1))$ 
decreases to 2.1.
According to the present results, the GT transitions from $^{12}$Be($0^+_3$) 
are weak, while the $B(GT:^{12}$Be($2^+_2)\rightarrow ^{12}$B($1^+_1))$
is largest among the $2^+$ states because the $2^+_2$ state is 
the other $0\hbar\omega$ state belonging to the $K^\pi=0^+_2$ band.

The present results concerning the weak $GT$ decay from the ground state
is consistent with the discussion given in Refs. \cite{BARKER,TSUZUKIa}.

\begin{table}
\caption{ \label{tab:be12beta} The strength of $\beta$ decays.
$B(GT)$ is defined as $|\langle \sigma\tau \rangle|^2$. The experimental
data are taken from \protect\cite{CHOU}.}
\begin{center}
\begin{tabular}{ccccc}
&initial & final & & \\
&$(J^\pi,E_x)$ (MeV) & $(J^\pi,E_x)$ (MeV) & B(GT) & \\
\hline
&   &&  exp. &\\ 
&$^{12}$Be($0^+$,0) & $^{12}$B($1^+$,0) & 0.56 &\\
\hline
\hline
& & & \multicolumn{2}{c}{theory} \\
& & &before diagonalization & after diagonalization\\
\hline
&$^{12}$Be($0^+_1$) & $^{12}$B($1^+_1$) & 0.2 &0.9 \\
&$^{12}$Be($0^+_2$) & $^{12}$B($1^+_1$) & 3.0 &2.1\\
&$^{12}$Be($0^+_3$) & $^{12}$B($1^+_1$) & 0.05 &0.07\\
&$^{12}$Be($2^+_1$) & $^{12}$B($1^+_1$) & 0.004 &0.1\\
&$^{12}$Be($2^+_2$) & $^{12}$B($1^+_1$) & 0.5 &0.4\\
\end{tabular}
\end{center}
\end{table}

\subsection{$E2$,$E1$,$E0$ transition strength}

Although the data concerning the $E2$ transition strength provide good
information about proton deformations, 
we must take care of the following points in the analysis of light 
neutron-rich nuclei.
Firstly, it is dangerous to assume the same deformation of 
the proton density as that of the neutron density in light unstable nuclei.
Secondly, the classical relations between the intrinsic deformation
($Q_0$) and the observables, which are often used in a simple analysis of 
heavy nuclei, do not necessarily work in light nuclei. 
Therefore, it is necessary to analyze the transition strength
based on a microscopic calculation.
As shown in previous AMD studies \cite{ENYObc,ENYOe,ENYOg},
the experimental $Q$-moments and $B(E2)$ values of light nuclei were
reproduced well by using bare charges,
because of the advantage of the AMD method, which 
can directly express the proton and neutron deformations.

Table \ref{tab:be12e2} gives the theoretical $B(E2)$, $B(E1)$ and
 $B(E0)$ values calculated by
the VAP calculation after diagonalization.
The intra-band $E2$ transitions in the $K^\pi=0^+_1$ band
are strong, as seen in $B(E2;2^+_1\rightarrow 0^+_1)$=14 e$^2$fm$^4$,
due to the deformed intrinsic state.
On the other hand, $B(E2)$ is smaller in the transition 
between the $2^+_2$ and $0^+_2$ states in the $K^\pi=0^+_2$ band, which have
rather spherical shapes compared to those in the $K^\pi=0^+_1$ band.
Although the deformations of the states in the 
$K^\pi=0^+_3$ band are extremely large in the present results,
$B(E2;4^+_2\rightarrow 2^+_4)$ and 
$B(E2;2^+_4\rightarrow 0^+_3)$ are not as large as 
those in the $K^\pi=0^+_1$ band,
because the intrinsic structure varies with increasing
the total spin ($J$) along the $0^+_3$ band.
The $B(E2)$ values regarding the transitions between 
different bands are smaller compared with those for the intra-band transitions.

Compared with the experimental data, $B(E2;2^+_1\rightarrow 0^+_1)$=
10.5$\pm$1.1 (e$^2$ fm$^4$) 
of $^{10}$Be, the present results predict a larger 
$B(E2)$ of $^{12}$Be as $B(E2;2^+_1\rightarrow 0^+_1)=$14 e$^2$fm$^4$.
We should point out that the $E2$ transition strength is sensitive
to the mixing ratio with the $0^+_2$ and $2^+_2$ states
as well as deformations. 
Since the $0^+_2$ and $2^+_2$ states are suggested to exist 
just above the $0^+_1$ and $2^+_1$ states, 
we should more carefully investigate the energy difference and  
the state mixing between the $K^\pi=0^+_1$ 
and $K^\pi=0^+_2$ bands before making conclusion.

The $E1$ transition strength was recently measured as 
$B(E1;0^+_1 \rightarrow 1^-_1)$ = 0.05 $e^2$fm$^2$ \cite{IWASAKI}, 
which is rather large compared with other light nuclei.
The calculated result, $B(E1;0^+_1\rightarrow 1^-_1)$=0.02 $e^2$fm$^2$,
reasonably agrees with this experimental data.
Before diagonalization, the deformed $0^+_1$ state possesses the 
strength of the $E1$ transition from the $1^-_1$ state, while the
transition to the spherical $0^+_2$ state is weak. 
After diagonalization, the strength in the $0^+_1$ state 
distributes in the $0^+_2$ state due to state mixing between
the $0^+_1$ and the $0^+_2$ states.
In other words, the rather large $B(E1;0^+_1 \rightarrow 1^-_1)$
is caused by a deformation of the ground state.
If the mixing of the $K^\pi=0^-$ component in the $1^-_1$ state is taken
into account as described in \ref{subsec:energy}, 
theoretical value of the $E1$ strength become large
as $B(E1;0^+_1\rightarrow 1^-_1)$ =0.14 $e^2$fm$^2$, because the
$K^\pi=0^-$ component has a similar intrinsic structure as that 
of the $0^+_1$ state. 

\begin{table}
\caption{\label{tab:be12e2} Theoretical results of 
$E2$, $E1$ and $E0$ transition strength, $B(E\lambda)$. 
$B(E0)$ is defined as $|<f|{1+\tau \over 2}r^2|i>|$. 
}

\begin{center}
\begin{tabular}{ccccc}
 transitions & band($K^\pi$) &  Mult. & present \\
\hline
$^{12}$Be;$2^+_1\rightarrow 0^+_1$ & in $0^+_1$ & 
 $E2$ & 14 (e$^2$ fm$^4$) \\
$^{12}$Be;$2^+_2\rightarrow 0^+_2$ & in $0^+_2$ & 
 $E2$ & 8 (e$^2$ fm$^4$)  \\
$^{12}$Be;$2^+_4\rightarrow 0^+_3$ & in $0^+_3$ &
 $E2$ & 8 (e$^2$ fm$^4$) \\
$^{12}$Be;$4^+_1\rightarrow 2^+_1$ & in $0^+_1$ & 
 $E2$ & 14 (e$^2$ fm$^4$) \\ 
$^{12}$Be;$4^+_2\rightarrow 2^+_4$ & in $0^+_3$ & 
 $E2$ & 10 (e$^2$ fm$^4$) \\
$^{12}$Be;$4^+_1\rightarrow 2^+_2$ & &
 $E2$ & 5 (e$^2$ fm$^4$) \\
$^{12}$Be;$4^+_2\rightarrow 2^+_1$ & &
 $E2$ & 5 (e$^2$ fm$^4$) \\
$^{12}$Be;$4^+_2\rightarrow 2^+_3$ & &
 $E2$ & 6 (e$^2$ fm$^4$) \\
\hline
$^{12}$Be;$0^+_1\rightarrow 1^-_1$ & &
 $E1$ & 2 $\times$ 10$^{-2}$ (e$^2$ fm$^2$) \\
$^{12}$Be;$0^+_2\rightarrow 1^-_1$ & &
 $E1$ & 2 $\times$ 10$^{-2}$ (e$^2$ fm$^2$) \\
$^{12}$Be;$0^+_2\rightarrow 0^+_1$ & &
 $E0$ & 1.7 (e fm$^2$) \\
\end{tabular}
\end{center}
\end{table}

\section{Discussions}\label{sec:discuss}

In the present calculations, many deformed rotational bands 
appeared in $^{12}$Be.
It was found that 2$\alpha$ core
is formed in most of the states as well as 
in other neutron-rich Be isotopes:$^{10}$Be and 
$^{11}$Be \cite{ENYOf,ENYOg,ENYObe11}. 
In this section,
we consider the intrinsic structures of $^{12}$Be while focusing on 
the clustering aspects and the behavior of the valence neutrons. 
The features of the intrinsic structures, 
the roles of valence neutrons and the inter-cluster motions
in the cluster states
are discussed. 

\subsection{Systematics of rotational bands}
By analyzing the structures of the intrinsic states, 
we can classify the ground and excited states into
rotational bands.
We superpose the wave functions, $P^{J\pm}_{MK}\Phi_{AMD}$,
projected from all of the intrinsic wave
functions, $\Phi_{AMD}({\bf Z}^{J_i,\pi_i}_{n_i})$,
obtained by VAP so as to diagonalize the Hamiltonian matrix. 
Although the final wave function $\Phi_{J^\pm_n}(^{12}{\rm Be})$ 
for the $J^\pm_n$ state after diagonalization is the linear combination of 
spin-parity projected AMD wave functions
$P^{J\pm}_{MK}\Phi_{AMD}({\bf Z}^{J_i,\pi_i}_{n_i})$,
we consider the AMD wave function
before diagonalization as being the intrinsic state
of the corresponding $J^\pm$ state, because
$P^{J\pm}_{MK}\Phi_{AMD}({\bf Z}^{J\pm}_{n})$
is found to be the major component of the final result 
$\Phi_{J^\pm_n}(^{12}{\rm Be})$ 
of the $J^\pm_n$ state, except for the 
$2^+_3$ and $2^+_4$ states.
In Table \ref{tab:be12over}, the squared amplitudes,
$\langle \Phi_{J^\pm_n}(^{12}{\rm Be})|
P^{J\pm}_{MK}\Phi_{AMD}({\bf Z}^{J\pm}_{n}\rangle^2$
are listed.
The amplitudes are larger than 0.6,
except for the $1^-_2$, $2^+_3$ and $2^+_4$ states.
In the negative-parity state of $1^-_2$,
the component of the original state
becomes less than 0.6 after diagonalization.
Considering that this state might be unstable, 
we neglect the $1^-_2$ level in the present results.
In the case of the $2^+_3$ state,
the main component of the $2^+_3$ state is not the projected AMD state,
$P^{2+}_{M0}\Phi_{AMD}({\bf Z}^{2+}_{3})$,
obtained by a VAP calculation for $J^\pm_n=2^+_3$, but is the
$P^{2+}_{M2}\Phi_{AMD}({\bf Z}^{3+}_{1})$, which is 
projected from the intrinsic state of the $3^+_1$ state.
This means that a new $2^+$ state appears in addition to
three $2^+$ states obtained by the VAP calculation, ${\bf Z}^{2+}_{1}$,
${\bf Z}^{2+}_{2}$  and ${\bf Z}^{2+}_{3}$.
The energy of this new $2^+$ state,
$P^{2+}_{M2}\Phi_{AMD}({\bf Z}^{3+}_{1})$,
 is slightly lower than that of
$P^{2+}_{M0}\Phi_{AMD}({\bf Z}^{2+}_{3})$.
As a result, the wave functions, 
$\Phi_{AMD}({\bf Z}^{3+}_{1})$ and $\Phi_{AMD}({\bf Z}^{2+}_{3})$, 
are regarded to be the intrinsic states of the $2^+_3$ and $2^+_4$ states,
respectively, although the amplitude of the original
VAP state, $P^{2+}_{M0}\Phi_{AMD}({\bf Z}^{2+}_{3})$, 
in the $2^+_4$ state is as small as 0.4 after diagonalization 
because of mixing the $2^+_3$ state with the $2^+_4$ state. 
The reason why the intrinsic state of the
$2^+_3$ state is not obtained in the VAP calculation 
with $(J^\pm, K')=(2^+, 0)$ is because the $2^+_3$
state belongs to the $K^\pi=2^+$ band.

\begin{table}
\caption{ \label{tab:be12over} The amplitudes of the corresponding 
VAP wave functions $P^{J\pm}_{MK}\Phi_{AMD}({\bf Z}^{J_i\pi_i}_{n_i})$
in the final wave functions ($\Phi_{J^\pm_n}(^{12}{\rm Be})$)
obtained after the diagonalization. 
Total intrinsic spins $\langle {\bf S}_p^2 \rangle$ 
and $\langle {\bf S}_n^2 \rangle$ for protons and neutrons
in the spin-parity projected states 
$P^{J\pm}_{MK}\Phi_{AMD}({\bf Z}^{J_i\pi_i}_{n_i})$ before the 
diagonalization are also listed. }
\begin{center}
\begin{tabular}{ccccc}
$J^\pm_n$ after diagonalization & 
$(J^{\pi_i},|K|,n_i)$ in VAP & 
$\langle \Phi_{J^\pm_n}(^{12}{\rm Be})|
P^{J\pm}_{MK}\Phi_{AMD}({\bf Z}^{J^{\pi_i}}_{n_i})\rangle^2$ 
& $\langle {\bf S}_p^2 \rangle$ & $\langle {\bf S}_n^2 \rangle$\\
\hline
$0^+_1$  &$(0^+,0,1)$ & 0.8 & 0.1 & 1.0\\
$2^+_1$  &$(2^+,0,1)$ & 0.8 & 0.1 & 1.1\\
$4^+_1$  &$(4^+,0,1)$ & 0.9 & 0.1 & 1.2\\
$6^+_1$  &$(6^+,0,1)$ & 0.9 & 0.0 & 1.5 \\
$8^+_1$  &$(8^+,0,1)$ & 0.9 & 0.1 & 1.6 \\
\hline
$0^+_2$  &$(0^+,0,2)$ & 0.6 & 0.2 & 0.1 \\
$2^+_2$  &$(2^+,0,2)$ & 0.7 & 0.9 & 0.1\\
\hline
$0^+_3$  &$(0^+,0,3)$ & 0.7 & 0.0 & 0.5\\
$2^+_4$  &$(2^+,0,3)$ & 0.4 & 0.0 & 1.0\\
$4^+_2$  &$(4^+,0,2)$ & 0.6 & 0.0 & 1.0\\
$6^+_2$  &$(6^+,0,2)$ & 0.6 & 0.0 & 1.2\\
\hline
$1^-_1$  &$(1^-,1,1)$ & 0.9 & 0.1 & 0.8\\
$2^-_1$  &$(2^-,1,1)$ & 0.9 & 0.1 & 0.8\\
$3^-_1$  &$(3^-,1,1)$ & 0.9 & 0.1 & 1.0\\
$4^-_1$  &$(4^-,1,1)$ & 0.7 & 0.1 & 1.1\\
$5^-_1$  &$(5^-,1,1)$ & 0.9 & 0.5 & 1.2\\
\hline
$2^+_3$  &$(3^+,2,1)$ & 0.8 & 0.0 & 2.1\\
$3^+_1$  &$(3^+,2,1)$ & 0.8 & 0.0 & 2.1\\
$5^+_1$  &$(5^+,2,1)$ & 0.8 & 0.0 & 2.0\\
$7^+_1$  &$(7^+,2,1)$ & 0.9 & 0.0 & 1.2\\
\hline
$0^-_1$  &$(0^-,0,1)$ & 1.0 & 0.1 & 1.7\\
$1^+_1$  &$(1^+,1,1)$ & 0.9 & 0.0 & 1.3\\
$1^+_2$  &$(1^+,0,1)$ & 1.0 & 1.9 & 0.3\\ 
$6^-_1$  &$(6^-,1,1)$ & 0.9 & 0.0 & 2.0 \\
\end{tabular}
\end{center}
\end{table}

By analyzing the intrinsic states, we find the rotational bands
$K^\pi=0^+_1, 0^+_2, 0^+_3, 2^+_1, 1^-_1$, which consist of 
the states 
$(0^+_1, 2^+_1, 4^+_1, 6^+_1, 8^+_1)$, 
$(0^+_2, 2^+_2)$,
$(0^+_3, 2^+_4, 4^+_2, 6^+_2)$
$(2^+_3, 3^+_1, 5^+_1, 7^+_1)$, 
and $(1^-_1, 2^-_1, 3^-_1, 4^-_1, 5^-_1)$, respectively.
We analyze the parity-eigen components in
the single-particle wave functions of the intrinsic states,
and find that each rotational band is 
dominated by either the $0\hbar\omega$, $1\hbar\omega$ or $2\hbar\omega$ 
excited configuration.
The states in the three positive-parity bands ($K^\pi=0^+_1, 0^+_3, 2^+_1$) 
have the dominant $2\hbar\omega$ excited configurations, while the states in 
the $K^\pi=0^+_2$ band are dominated by 
the $2\hbar\omega$ configurations.
The first point is that the states in the ground band are not 
the neutron shell-closed states, but are the 
prolately deformed intruder states 
with developed clustering, even though this
$^{12}$Be nucleus has a neutron magic number of $N=8$. 
This is consistent with discussions of the vanishing of the 
neutron magic number in $^{12}$Be given by Refs.
\cite{BARKER,FORTUNE,ITAGAKIa,IWASAKI,TSUZUKIa}.
The $K^\pi=0^+_1$ band 
starting from the $0^+_1$ state terminates with the $8^+_1$ state,
because the $J^\pi=8^+$ is the highest spin in the 
$2\hbar\omega$ configurations.
On the other hand,
the present results predict that the $0^+_2$ and $2^+_2$ states with 
the neutron $p$-shell closed structures constitute 
the excited band, $K^\pi=0^+_2$, just above the ground band.
The second interesting point is 
that the newly observed levels, $4^+$ and $6^+$ states \cite{FREER},
are considered to belong to 
the higher band of the two bands($K^\pi=0^+_1$ and $K^\pi=0^+_3$),
 both of which 
are the $2\hbar\omega$ excited states with developed cluster structures.
The $K^\pi=2^+_1$ band 
consists of the $2^+_3,3^+_1,5^+_1,7^+_1$ states, which have 
the other $2\hbar\omega$ configurations .
Although spin-parity eigen states, $4^+$ and $6^+$, 
are able to be projected from the intrinsic states of this
$K^\pi=2^+_1$ band in 
principle, the $4^+$ and $6^+$ states in the $K^\pi=2^+_1$ band 
can not be identified, because these states strongly mix with the 
$4^+_1$, $4^+_2$, $6^+_1$ and $6^+_2$ states in the $K^\pi=0^+_1$ and
$K^\pi=0^+_3$ bands.
We find a negative-parity band, $K^\pi=1^-_1$,
in which the states($1^-$,$2^-$,$3^-$,$4^-$,$5^-$) 
are dominated by $1\hbar\omega$ configurations.

\subsection{Intrinsic structures}
\begin{figure}
\noindent
\epsfxsize=0.4\textwidth
\centerline{\epsffile{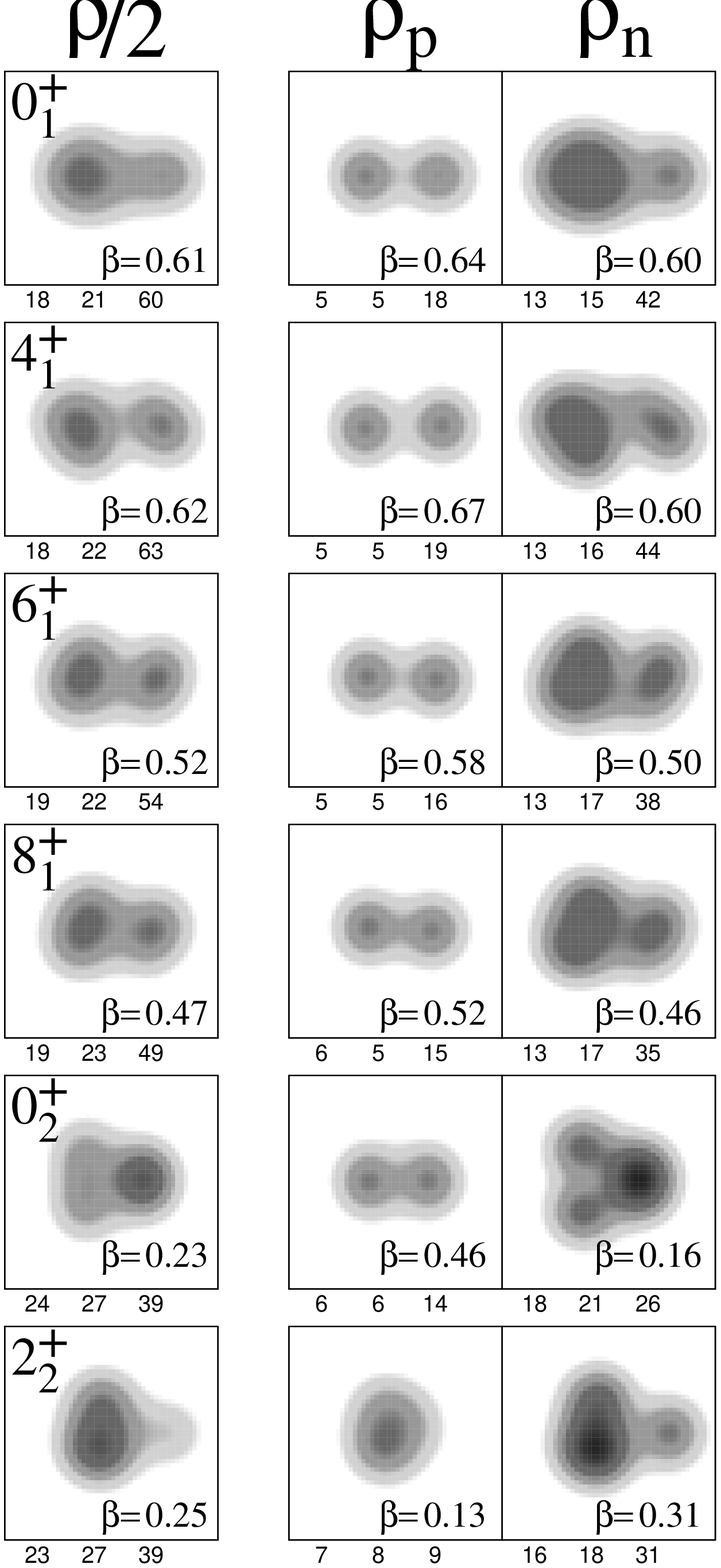}\qquad\qquad \ \ 
\epsfxsize=0.4\textwidth
\epsffile{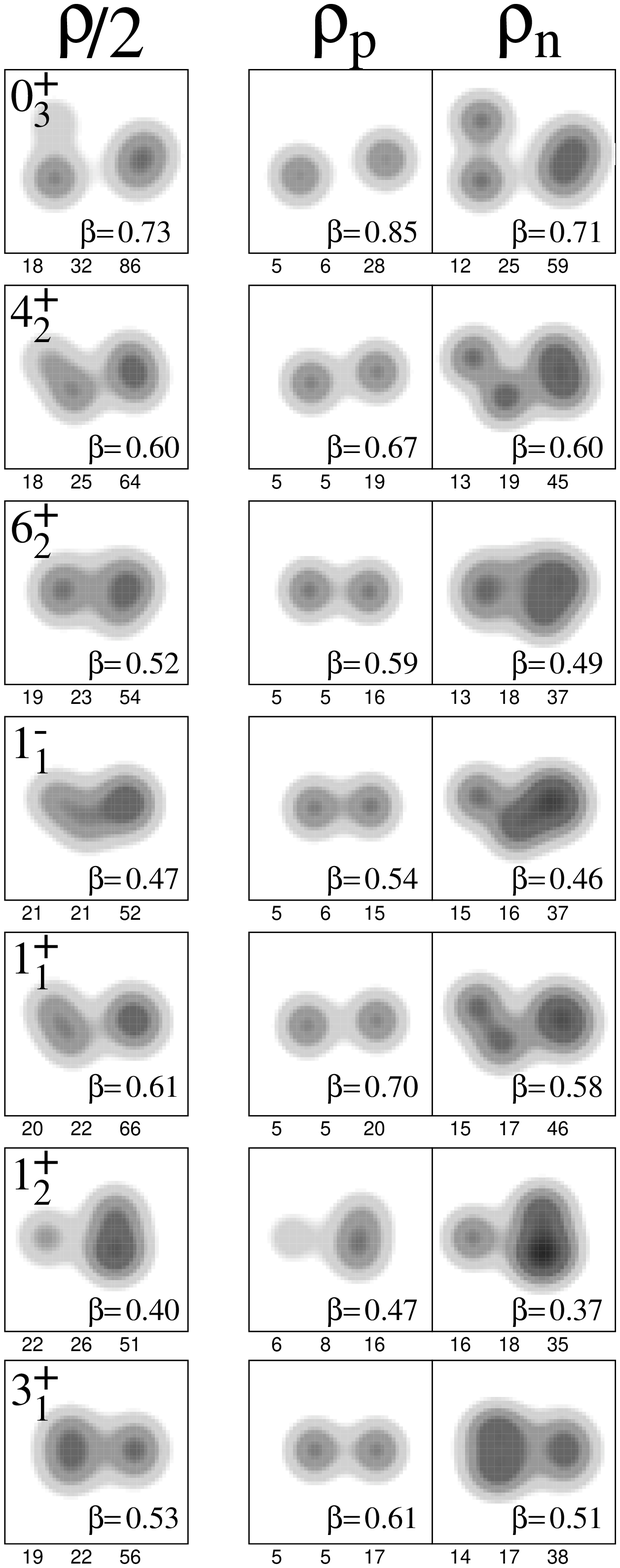}}
\caption{\label{fig:be12dens}
Density distribution of the intrinsic state 
$\Phi_{AMD}({\bf Z}^{J\pm}_n)$ before projection and
diagonalization. The intrinsic system is projected 
onto a plane which contains the approximately longitudinal axis of the 
intrinsic state.
The density is integrated along a transverse axis perpendicular to the
plane.
The densities for matter, protons and neutrons are 
presented at the left, middle 
and right, respectively. 
The deformation parameters $\beta$ are written at the bottom of 
the figures.
The expectation values of $xx$, $yy$, $zz$
for the matter, proton and neutron densities are written below 
the figures. The size of the frame box is 10 fm$\times$10 fm. 
}
\end{figure}

We discuss the features of the intrinsic structures, such 
as the deformations and clustering aspects.
The density distributions of the intrinsic AMD wave functions 
are shown in Fig. \ref{fig:be12dens}.

In most of the states,
except for the $2^+_2$, $5^-_1$ and $1^+_2$ states,
the densities of protons have dumbbell-like shapes, which 
indicate the formation of a 2$\alpha$ core.
On the other hand, 2$\alpha$ core breaking occurs in 
the $2^+_2$, $5^-_1$ and $1^+_2$ states.
For the quantitative discussions, we can estimate the degree of  
core breaking based on the non-zero values of the 
squared total-intrinsic spin of protons, 
$\langle {\bf S}_p^2 \rangle$, given in 
table \ref{tab:be12over}. In most of the states, 
the values $\langle {\bf S}_p^2 \rangle$ almost equal to 0, 
which indicates only a slight breaking of the 2$\alpha$ core.
However, as seen in the large $\langle {\bf S}_p^2 \rangle$ values
in the $2^+_2$ and  $5^-_1$ states, the 2$\alpha$ core structures 
somehow dissociate in these band terminal states of the 
$K^\pi=0^+_2$ and $1^-$ bands because of the spin-alignment effect.
On the other hand, the $1^+_2$ state has 
a quite different structure from that of the 2$\alpha$-core state.
It is a spin-aligned state with proton spin, ${\bf S}_p\sim 1$, where
one of the 2-$\alpha$ clusters is completely broken. 
In fact, no dumbbell shape is seen 
in the proton density of this state (Fig. \ref{fig:be12dens}).
The alignment of the intrinsic spins is the  
origin of the unnatural spin parity of this 
state. This $1^+_2$ state is analogous to the  
$3/2^-$ state at 8.4 MeV in 
$^{11}$Be, and also to the $1^+_1$ state 
in $^{12}$C, in which cluster breaking occurs due to the 
the aligned intrinsic spins.

The deformation parameter ($\beta$) of the proton density shown in 
Fig. \ref{fig:be12dens} is a useful quantity
in order to quantitatively discuss the developments of clustering 
in the 2$\alpha$ core states.
Here, we explain the cluster developments in relation to the neutron 
structure, such as the neutron deformation and neutron intrinsic 
spin.

In the $K^\pi=0^+_1$ band, the states have a prolate deformation 
with the developed 2$\alpha$ core, as shown in Fig. \ref{fig:be12dens}. 
In the $0^+_1$, $2^+_1$ and $4^+_1$ states, the 2$\alpha$ core 
well develops following prolate deformation of the neutron density, 
which is caused by the 2 neutrons in a longitudinal $sd$-orbit.
The ground $0^+_1$ state has an approximately axial symmetric shape 
(Fig. \ref{fig:be12dens}), which is considered to be caused by
the symmetric neutron configuration described by a $p_{3/2}$ sub-shell 
closure and the axial symmetric $sd$-orbit with $(\lambda\mu)=(20)$ symmetry
in the $SU_3$ limit.
The non-zero total intrinsic spin of neutrons
of the ground $0^+_1$ state shown in 
Table \ref{tab:be12over} is considered to come 
from a sub-shell closure effect of the neutron $p_{3/2}$-shell.
As the total spin increases, the alignment of the intrinsic spins
of the $sd$-shell neutrons further increases the total intrinsic 
spin of neutrons(Table \ref{tab:be12over}).
The aligned $sd$-shell neutrons make the neutron-density deformation 
of the nucleus to be smaller in the high-spin region, $J\ge 6$. 
As a result, the clustering become weak in the $6^+_1$ and $8^+_1$ states,
as found in the reduction of the proton deformation parameter
(Fig. \ref{fig:be12dens}).

In the $K^\pi=0^+_2$ band with the dominant $0\hbar\omega$ configurations, 
the neutron deformation is smaller than that in the $K^\pi=0^+_1$ band.
In the $0^+_2$ state, 4 neutron pairs make a small
tetrahedron-shape structure, which is approximately equivalent to the $p$-shell
closed structure. Because of the shell effect of the neutron $p$-shell
closure, the clustering  is smaller than that in the $0^+_1$ state.
Although the intrinsic state of the band terminal $2^+_2$ state in this band 
has the prolate neutron density as seen in Fig. \ref{fig:be12dens},
the spin-parity $J^\pm=2^+$ eigen state projected from this intrinsic state
is dominated by the neutron $p$-shell closure component.
The total spin $J=2$ in this $2^+_2$ state is composed of the 
aligned total-angular momentum of the $p$-shell protons. 
The overlap 
$|\langle P^{2+}_{MK=0}\Phi_{AMD}({\bf Z}^{0+}_2)|
P^{2+}_{MK=0}\Phi_{AMD}({\bf Z}^{2+}_2)\rangle|^2 $
is still large as about 0.6.  
The $j$-$j$ coupling feature in this band terminal state is found 
in the non-zero value of the 
squared total-intrinsic spin of protons(Table \ref{tab:be12over}) 
and also in the disappearance of the dumbbell-like shape of proton 
density(Fig. \ref{fig:be12dens}).

In the $K^\pi=0^+_3$ band, the $0^+_3$ and $2^+_4$ states have extremely
developed cluster structures, like $^6$He+$^6$He. One of the reasons 
for the development of clustering is considered to be the orthogonal
condition to the $0^+_1$ and $2^+_1$ states in the lower band.
The inter-cluster distance shrinks with increasing total spin, 
due to the spin alignment.
In the $0^+_3$, $2^+_4$ and $4^+_2$, low matter density regions 
between the clusters are seen (Fig. \ref{fig:be12dens}), while they 
disappear in the $6^+_2$ state.
The intrinsic structure of the $K^\pi=0^+_3$ band changes 
rather rapidly with an increase of
the total spin, as can be seen in the decrease of the 
deformation parameter $\beta$ (Fig. \ref{fig:be12dens}) and also 
in the spin alignment of the neutron intrinsic spins 
(Table \ref{tab:be12over}). Eventhough the $^6$He+$^6$He cluster structure 
changes due to the spin alignment in this $K^\pi=0^+_3$ band, 
it is help to consider the spatial symmetry in the $SU_3$ representation. 
Since the intrinsic state of the band head $0^+_3$ has 
$(\lambda\mu)$=(24) symmetry in the $SU_3$ limit, 
which provides a $K=0$ band from 
$J^\pi=0^+ \rightarrow 6^+$, it is reasonable that $K^\pi=0^+_3$ band
terminates at the $6^+_2$ state. 

In spite of the mixing of the states and the spin alignments, 
all of the states in the $K^\pi=0^+_1$, $0^+_3$ and $2^+_1$ 
bands are dominated by $2\hbar\omega$
configurations with 2 neutrons in the $sd$-shell.
It is another interesting problem to search for the weak-coupling
cluster structure, which is described by 
the relative motion between the 2-$^6$He clusters. 
In pioneering work by M. Ito et al.\cite{ITO}, 
weak coupling states with 2-$^6$He clusters in $^{12}$Be were studied.
Although the calculation was not fully microscopic, because
they did not make antisymmetrization of 
the neutrons between clusters, it is interesting that 
He-cluster states were suggested to appear near and above the 
threshold energies.
The presently predicted $0^+_3$ state is a candidate
of the weak coupling cluster state, 
 because of the large relative 
distance between clusters. 
However, in the $4^+_2$ and $6^+_2$ states in the $K^\pi=0^+_3$ bands,
the components of the weak coupling cluster state may not be
large, since the He-cluster structure becomes weak due to the spin 
alignments. 
Instead, it is expected that 
the molecular resonances with a weak-coupling cluster 
structure may exist above those spin-aligned states.
We performed a simple VAP calculation of the higher states,  and
found that candidates of the weak-coupling states 
appear above those spin-aligned states with $2\hbar\omega$ configurations
obtained in the present calculations.
We find a $4^+_3$ state with a developed $^6$He+$^6$He-cluster structure 
at a 0.6 MeV higher energy than the $4^+_2$ state.  
Although the $6^+_3$ state was unstable for cluster escaping in the present 
VAP calculations using $m=0.65$,
we obtained a $6^+_3$ state with a developed $^6$He+$^6$He structure
 at a few MeV higher energy than the $6^+_2$ state
by using $m=0.62$.
Such weakly bound states should be carefully investigated 
by taking care of the threshold energies and the 
stability for particle decays.

The negative-parity states $1^-_1$, $2^-_1$, $3^-_1$, $4^-_1$
and $5^-_1$ in the $K^\pi=1^-_1$ band are deformed states with 
cluster structures. The deformed structures of the neutron density
are made from the dominant $1\hbar\omega$ excited configurations with one 
neutron in the $sd$-shell. The degree of prolate deformation of the 
neutron density in this band 
is smaller than the states in the $K^\pi=0^+_1$ band. 
The smaller neutron deformation causes the weaker cluster development
in the $K^\pi=1^-_1$ band compared with those in the $K^\pi=0^+_1$ band.
The $1^-_1$, $2^-_1$, $3^-_1$, $4^-_1$ states have axial-asymmetric 
neutron structures, while the band terminal $5^-$ state has  
approximately axial-symmetric neutron density.
The reason for $K^\pi=1^-$ of the $1^-_1$ state 
can be naturally understood by a weak coupling picture of $^{11}$Be core 
and a neutron. In the simple weak coupling picture, since
the $1^-_1$ state of $^{12}$Be can be described by the coupling of 
a $^{11}$Be($1/2^-$) core with ($\lambda\mu$)=(21) $SU_3$ symmetry 
and a $s_{1/2}$-orbit neutron, the $K^\pi=1^-$ in the $^{12}$Be($1^-$) state
come from the $K^\pi=1^-$ in the $^{11}$Be core. 
In the recent work with a three-body(n+n+$^{10}$Be) model \cite{NUNES}, 
in which the low-lying states of $^{12}$Be were studied 
from the weak coupling picture, the excitation energy of the $1^-_1$ state 
was predicted to be about 3 MeV.
Another possible negative parity band is $K^\pi=0^-$, which is a parity
doublet band of a $^8$He+$^4$He cluster structure,  
as is suggested in a He+He cluster model by P. Descouvement et al. 
\cite{DESCOUVEMENT}.
We found a $1^-$ state with the $^8$He+$\alpha$ cluster structure 
by the VAP calculation for $K^\pi=0^-$ at slightly higher 
energy than that for $K^\pi=1^-$.
Even though the mixing of this $K^\pi=0^-$ component is important to lower 
the excitation energy of the $1^-_1$ state as mentioned before 
in \ref{sec:results}, the major component of the $1^-_1$ state is 
the $K^\pi=1^-$ state. The improved results including the $K^\pi=0^-$ 
configuration suggest the higher state($1^-_2$) with the significant 
$K^\pi=0^-$ component may appear at 9 MeV excitation energy, 
which is consistent with the prediction in Ref.\cite{DESCOUVEMENT}.

Next, we discuss the structures of the unnatural spin-parity states.
As mentioned above, the unnatural spin-parity of the $1^+_2$ state 
is caused by the aligned intrinsic spins of protons. On the other hand,
the unnatural spin-parity of the other states ($3^+_1$, $5^+_1$, $7^+_1$, 
$1^+_1$ and $0^-_1$) originates from the aligned intrinsic spins of the 
$sd$-shell neutrons. 
Alignments of the neutron intrinsic spins were found 
in the $2^+_3$, $3^+_1$ and  $5^+_1$ states, as seen
in the expectation values of the squared total neutron spins, 
$\langle {\bf S}^2_n \rangle \sim 2$, presented in Table \ref{tab:be12over}.
The spin alignment is mainly caused by 2
$sd$-shell neutrons.
The deformation and the density distribution of the intrinsic 
state of the $7^+_1$ state are very similar to those of the 
$3^+_1$ and $5^+_1$,
though $\langle {\bf S}_n^2 \rangle$ in the $7^+_1$ state
is slightly smaller.
The $1^+_1$ and $0^-_1$ states with unnatural spin-parity have  
aligned intrinsic spins of neutrons, like the states 
in the $K^\pi=2^+_1$ band.
Although a $^6$He+$^6$He-like structure is found in the $1^+_1$ state,
one of the $^6$He clusters is an excited $^6$He with ${S}_n=1$ components.

Here, we stress that it is essential to consider the orthogonality to the
lower states as well as antisymmetrization 
of all the nucleons in a study of the excited states 
of $^{12}$Be, because the present results 
suggest the existence of many 2$\alpha$ core states
in the low-energy region. Although these states 
are not necessarily written in terms of $^6$He+$^6$He clusters,
such lower states must have effects on the higher weak-coupling cluster 
states.
The present results indicate the importance of treating degrees of 
freedom of the intrinsic spins for
all the nucleons in a study of $^{12}$Be. 
The reasons can be summarized as follows.
First of all, a closed $p_{3/2}$-shell of neutrons plays an important role for 
the energy gain of the intruder ground state. 
Secondly, the alignments of intrinsic spins as
$S=1$ are necessary to describe the unnatural 
spin-parity states. In the third point,
the spin alignment of the neutron intrinsic spins occurs in high spin states.

\subsection{Picture of molecular orbits}

As mentioned in the previous subsection, 
the 2$\alpha$ core structure appears in many 
of the states. The development of clustering is sensitive 
to the prolate deformation of the neutron density. We inspected the neutron 
structure based on the single-particle behavior of the valence 
neutrons surrounding the 2$\alpha$ core. We also considered the relation
between the neutron orbits and the cluster development.
We analyzed the single-particle wave functions of the 
valence neutrons in the intrinsic states.
The detailed formulation of how to extract 
the single-particle energies and wave functions from an AMD wave function 
is described in Refs. \cite{DOTE,ENYOg}.

According to the analyses by the single-particle orbits and the
single-particle energies,
the 2$\alpha$ core is composed of nucleons occupying the lowest
4 proton orbits and 4 neutron orbits.
The higher 4 neutron orbits correspond to those of the valence neutrons
surrounding the 2$\alpha$ core.
By extracting the positive and negative components in the orbits,
many of the single-particle orbits are found to be
approximately parity-eigen states, as the amplitude of the 
dominant parity-eigen state is more than 70\% in each orbit.
Since the negative and positive-parity orbits of 
the valence neutrons are associated with the $p$-orbits and $sd$-orbits,
respectively, we can classify the ground and excited states of 
$^{12}$Be in terms of the $n\hbar\omega$ excitation, where $n$ 
is the number of the valence neutrons with the dominant positive-parity 
components. The $K^\pi=0^+_2$ band is approximately described 
by $0\hbar\omega$ configurations,
while the $K^\pi=0^+_1$, $0^+_3$, $2^+_1$ bands are dominated
by $2\hbar\omega$ configurations with 2 neutrons in $sd$-like orbits.
On the other hand, the main components of 
the $K^\pi=1^-_1$ band are $1\hbar\omega$ configurations.

\begin{figure}
\noindent
\epsfxsize=0.49\textwidth
\centerline{\epsffile{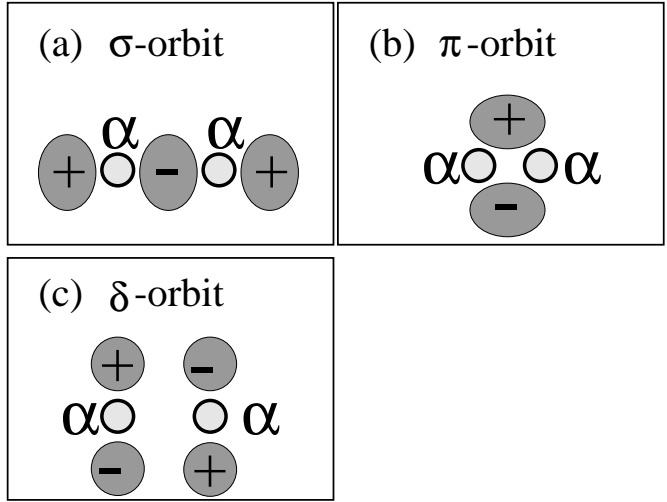}}
\caption{\label{fig:sigmapi}
Sketches for the molecular orbits, (a) $\sigma$-orbits, (b)$\pi$-orbits,
and (c)$\delta'$-orbits surrounding 2$\alpha$ core. 
These molecular orbits are explained by linear
combinations of the $p$-shell orbits around the $\alpha$ cores.
}
\end{figure}

The idea of molecular orbits surrounding a 2$\alpha$ core
is helpful to understand
the roles of the valence neutrons in neutron-rich Be isotopes.
The molecular orbits in Be isotopes were suggested in a study of $^9$Be
with a $2\alpha+n$ cluster model \cite{OKABE}. 
They assumed $\sigma$-orbits and $\pi$-orbits which are made from 
linear combinations of the $p$-orbits around the $\alpha$ cores
(see Fig.\ref{fig:sigmapi}).
This idea was applied to neutron-rich Be isotopes by Seya et al. a long time
ago \cite{SEYA}. In the 1990's von Oertzen et al. \cite{OERTZEN,OERTZENa}
revived this kind of research 
to understand the rotational bands of neutron-rich Be isotopes, and Itagaki 
et al. \cite{ITAGAKI,ITAGAKIa} described the structures 
of the low-lying states of $^{10}$Be and $^{12}$Be
by assuming 2$\alpha$ core and valence
neutrons in the molecular orbits. 
The formation of the 2$\alpha$ and valence neutron structures in neutron-rich 
Be isotopes was first guaranteed theoretically by the AMD
calculation \cite{ENYObc,DOTE,ENYOf,ENYOg,ENYObe11}, 
where the existence of any clusters or molecular orbits was not assumed.
In these AMD studies, the viewpoint of the 
molecular orbit was found to be useful to understand 
the cluster development in $^{10}$Be and $^{11}$Be.
Therefore, it is an interesting problem whether the states of $^{12}$Be can be
described by the molecular orbits.

\begin{figure}
\noindent
\epsfxsize=0.49\textwidth
\centerline{\epsffile{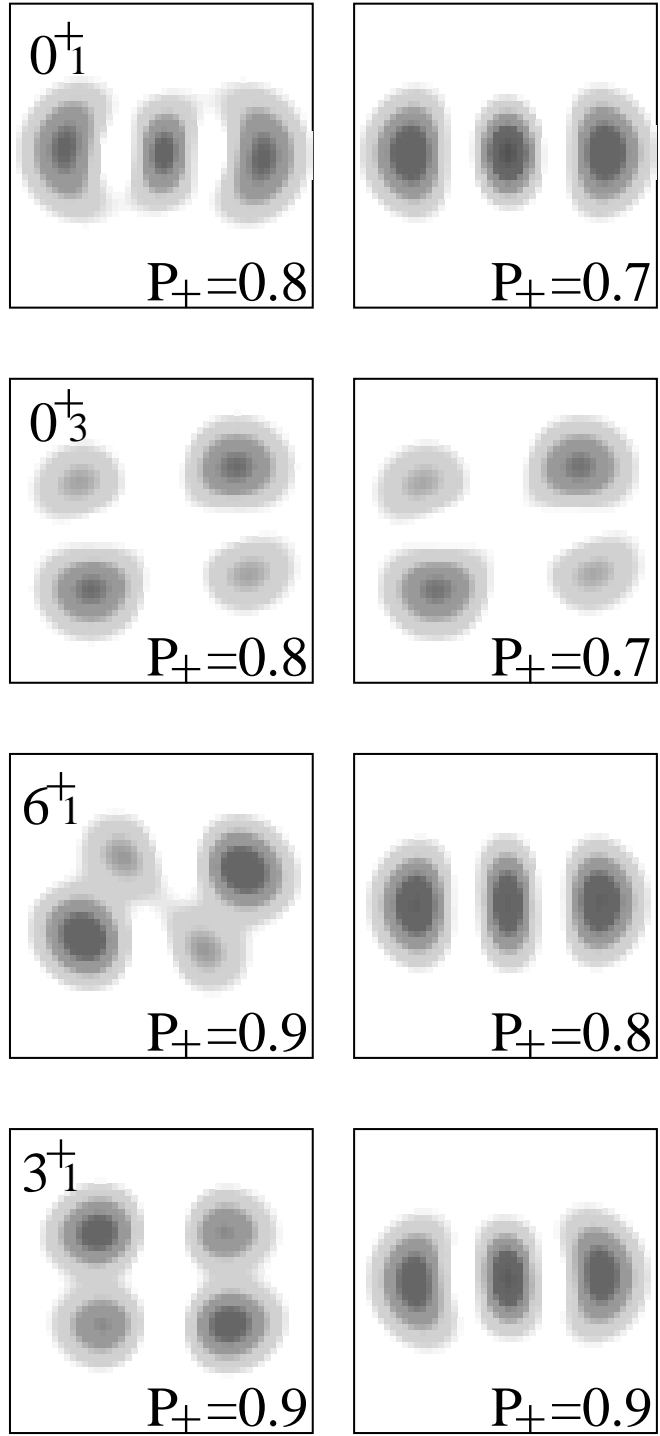}}
\caption{\label{fig:singles}
Density distributions of the single-particle wave functions of the 
valence neutrons in the 
intrinsic wave functions of the 
$0^+_1$, $0^+_3$, $6^+_1$ and $3^+_1$ states.
The figures at the left(right) show the densities regarding 
the positive-parity 
components of the first(second) highest neutron orbits. 
The value $P_+$ in each orbit indicates the squared amplitude of the 
contained positive-parity component.}
\end{figure}

In the present results for $^{12}$Be, 
we find a new kind of molecular orbit
besides the suggested $\pi$-orbit and $\sigma$-orbit.
In the positive-parity orbits of the valence neutrons in $^{12}$Be,
two kinds of molecular orbits appear, both of which are associated 
with $sd$-orbits.
The first one is the $\sigma$-orbit (Fig. \ref{fig:sigmapi}a), 
while the second one is
a quite new molecular orbit, shown in Fig. \ref{fig:sigmapi}c.
This orbit is the other positive-parity orbit 
made from a linear combination of the 
$p$-orbits around the $\alpha$ cores. 
As shown in Fig. \ref{fig:sigmapi}c, the combined $p$-orbits in 
this orbit are perpendicular to those in the $\sigma$-orbit.
We call this new positive-parity orbit 
a $\delta'$-orbit in the present paper, although it has $(\lambda \mu)=(01)$
symmetry in the $SU_3$ limit, which is 
perpendicular to the so-called
$\delta$-orbit in the field of the molecular physics. 
In the case of $^{12}$Be, the negative-parity orbit of the 
neutron surrounding 2$\alpha$ does not necessarily correspond 
to the pure molecular $\pi$-orbit,
because the $p_{3/2}$-shell closure can not be
described by simple $\pi$-orbits,
Therefore, in the following discussions, we concentrate on 
the positive-parity orbits of the valence neutrons associated
with the molecular $\sigma$-orbits and $\delta'$-orbits.

Fig. \ref{fig:singles} shows the 
density distributions of the single-particle wave functions of the 
first and second highest neutron orbits. 
In the low-spin cluster states,
the positive-parity orbits of the valence neutrons 
can be well associated with the two types of the molecular orbits
($\sigma$ and $\delta'$).
In the $0^+_1$ state, two valence neutrons with up and down spins
occupy the 
$\sigma$-like orbits, which have 2 nodes along the longitudinal axis. 
In the $0^+_3$ state, which is dominated by the other 
$2\hbar\omega$ configurations, the two neutrons occupy $\delta'$-like orbits.
 It is very surprising that the developed $^6$He+$^6$He
cluster structure in the $0^+_3$ state is understood by the new 
molecular $\delta'$-orbits. It occurs when 2 deformed $^6$He clusters
are attached in parallel.
In the $0^+_2$ state, all of the 4 valence neutrons are in the 
negative-parity orbits.
Comparing the energies of the $0^+_3$ state with those of the
$0^+_1$ and $0^+_2$ states, the $\delta'$-orbit is the 
highest among the molecular orbits ($\sigma$, $\pi$ and $\delta'$).

The molecular $\sigma$-orbit is one of the reasons for the 
deformed ground state of $^{12}$Be with the $2\hbar\omega$ 
configurations, which is lower than the closed neutron-shell 
state.
Since Be nuclei prefer
prolate deformations because of the 2$\alpha$-cluster core,
the $\sigma$-orbit gains kinetic energy in the
developed cluster system.
In pioneering studies, \cite{ITAGAKIa,ENYObe11},
the importance of the $\sigma$-orbit in the ground states of 
$^{11}$Be and $^{12}$Be were discussed in relation to a
vanishing of the magic number. 
Thus, the neutrons in the $\sigma$-orbit play an 
important role in the cluster development of the ground state of $^{12}$Be.
On the other hand, in case of the $0^+_3$ state,
we consider that the cluster development is further enhanced
due to the effect of the orthogonal condition to the $0^+_1$ state.

The next problem is whether or not the molecular orbits appear in the
unnatural spin-parity $3^+_1$ state, which has another deformed structure
dominated by the $2\hbar\omega$ configuration.
According to an analysis of single-particle wave functions,
as shown in Fig. \ref{fig:singles},
we find that the positive-parity orbits of the $3^+_1$ state consist of a
$\sigma$-like orbit and a $\delta'$-like orbit.
As mentioned before, the total intrinsic spin
of neutrons equals to that in the $3^+_1$ state (see Table \ref{tab:be12over}).
This means that the intrinsic spins of two neutrons in the 
$\sigma$-orbit and the $\delta'$-orbit are aligned to be one.
It is a unique character of the $3^+_1$ state to be different from 
the other $2\hbar\omega$ bands where
the intrinsic spins of two neutrons in the 
same spatial positive-parity orbits couple off to be zero. 

As mentioned above, in the case of the low-spin states in the deformed band, 
the highest two positive-parity orbits of the valence neutrons 
can be well associated with the $\sigma$ and $\delta'$-orbits.
However the positive-parity orbits in the high-spin states 
can not be classified by the simple $\sigma$ and $\delta'$-orbits,
but are related to the mixed orbits of the two.
For example, the 
highest neutron orbit in the $6^+_1$ states indicates the mixing of 
$\sigma$ and $\delta'$, as shown in Fig. \ref{fig:singles}.
As the total spin increases from $0^+_1$ to $6^+_1$ in the $K^\pi=0^+_1$
band, the alignment of the neutron intrinsic spins grows 
accompanying mixing the $\delta'$-orbit in the $\sigma$-orbit.

The states in the negative-parity band($K^\pi=1^-_1$) are 
described by $1\hbar\omega$ configurations, because one of the highest 2 
orbits is dominated by the positive-parity component, 
while the other one is almost a negative-parity orbit. 
The interesting point is that
the 30\% mixing of the positive-parity component in
the negative-parity orbit gives rise to the axial
asymmetric shape of the intrinsic state, as shown by the neutron density 
(Fig. \ref{fig:be12dens}).

In an analysis of $^{12}$Be with the molecular orbits, it should be 
stressed that all of the states with a 2$\alpha$ core can not be necessarily
described by pure molecular 
orbits: $\pi$, $\sigma$ and $\delta'$-orbits. Firstly, the 
single-particle orbits of the valence neutrons are not 
parity-eigen orbits. For example, 
although the highest two neutron orbits in the $0^+_1$ state
are dominated by positive-parity components, they 
contain 20\%$\sim$30\% mixing of negative-parity contamination,
which can not be regarded as the $\pi$-orbits but includes higher-shell 
configurations.
Moreover, mixing of the $\sigma$ and $\delta'$-orbits is 
found in the high-spin 
states in the $K^\pi=0^+_1$ and  $K^\pi=0^+_3$ bands, as can be seen in the
highest single-particle orbits of the $6^+_1$ state (Fig. \ref{fig:singles}).

\subsection{Systematics of the 2$\alpha$ clustering in Be isotopes}

\begin{figure}
\noindent
\epsfxsize=0.4\textwidth
\centerline{\epsffile{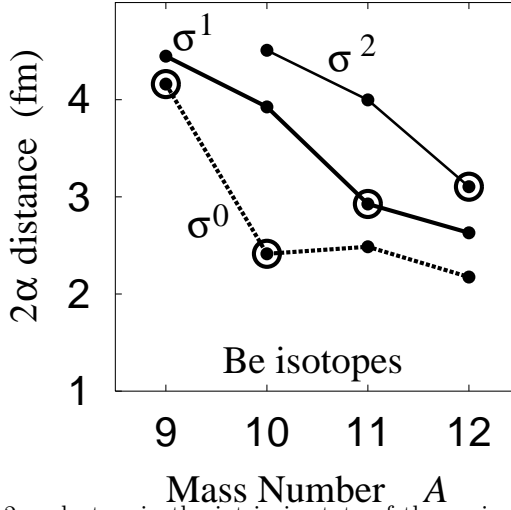}}
\caption{\label{fig:be-rpp}
Relative distance between 2-$\alpha$ clusters in the intrinsic state of the
major component described by a single AMD wave function.
The distance $d$ is evaluated by the relative distance between the 
mean centers of 2 proton pairs, defined as 
$d \equiv \protect\sqrt
({\rm Re}
[{\bf Z}_1+{\bf Z}_2-{\bf Z}_3-{\bf Z}_4])^2/2\protect\sqrt{\nu}$.
The dotted, thick solid and thin solid lines correspond to the distances in the
band head states with zero($\sigma^0$),
one($\sigma^1$) and two($\sigma^2$) neutrons in the $\sigma$ orbits.
Namely, the lines ($\sigma^0$,$\sigma^1$ and $\sigma^2$) correspond to 
\{$^9$Be($1/2^-_1$), $^{10}$Be($0^+_1)$, 
$^{11}$Be($1/2^-_1$), $^{12}$Be($0^+_2$)\}, 
\{$^9$Be($1/2^+_1$), $^{10}$Be($1^-_1)$, 
$^{11}$Be($1/2^+_1$), $^{12}$Be($1^-_1$)\} and 
\{$^{10}$Be($0^+_2)$, 
$^{11}$Be($3/2^-_2$), $^{12}$Be($0^+_1$)\}, respectively.
The open circles indicate the ground states. 
}
\end{figure}

 When we roughly regard the valence neutron orbits 
in the $0^+_1, 1^-_1, 0^+_2$ states of $^{12}$Be 
as the $\pi$-orbits and the $\sigma$-orbits,
these $^{12}$Be states remind us of analogous states 
in $^9$Be, $^{10}$Be and $^{11}$Be. 
In terms of the molecular orbits surrounding the 2$\alpha$ core,
the typical rotational bands in these Be isotopes are classified by
the number of occupied $\sigma$-orbits. Here, we use the notation
$\pi^m\sigma^n$ for those states having
$m$ neutrons in the $\pi$-orbits and $n$ neutrons in the $\sigma$-orbits
around a 2$\alpha$ core.
The valence neutron orbits  
in the lowest natural parity states($^9$Be($1/2^-$), $^{10}$Be($0^+_1)$, 
$^{11}$Be($1/2^-$) and $^{12}$Be($0^+_2$)) are described as
$\pi^m\sigma^0$, and those in the lowest unnatural parity states
($^9$Be($1/2^+$), $^{10}$Be($1^-)$, 
$^{11}$Be($1/2^+$) and $^{12}$Be($1^-$)) are noted as 
$\pi^{m-1}\sigma^1$,
while the $^{10}$Be($0^+_2)$, 
$^{11}$Be($3/2^-_2$) and $^{12}$Be($0^+_1$) states correspond to
$\pi^{m-2}\sigma^2$, where $m$ is the number of valence neutrons around
the 2-$\alpha$ core (see also Refs. \cite{OERTZEN,ENYOg,ENYObe11,OKABE}).
It is important that the number of neutrons in the $\sigma$-orbits 
influences the spatial development of the 2$\alpha$ core.
The changes in the relative distance between the 2 $\alpha$ clusters
in the Be isotopes are shown in Fig. \ref{fig:be-rpp}.
The dotted, thick solid and thin solid lines correspond to the distances in the
band head states with zero ($\pi^m\sigma^0$),
one ($\pi^{m-1}\sigma^1$) and two ($\pi^{m-2}\sigma^2$) neutrons 
in the $\sigma$-orbit.
In each Be isotope, 
the 2-$\alpha$ distance is smallest in the $\sigma^0$ state and
becomes larger and larger with increasing number of the 
occupied $\sigma$-orbits,
because of the energy gain of the $\sigma$-orbits in the developed cluster
system. Comparing the  2-$\alpha$ distance in each line, the dependence 
on the neutron number shows that the  
cluster development systematically decreases 
as the number of the occupied $\pi$-orbits increases. 
Especially, a drastic change in the $\sigma^0$ states 
from $^9$Be to $^{10}$Be is caused by the $p_{3/2}$ sub-shell closure
effect.
Also, in the $\sigma^1$ and $\sigma^2$ lines,
there exist gaps between the states with a closed $p_{3/2}$-shell
and those with an open
$p_{3/2}$-shell.
Thus, the closed $p_{3/2}$-shell plays an important role in weakening
the $2\alpha$ development.

One of the interesting features of neutron-rich 
Be isotopes is vanishing of the neutron magic number $N=8$ in $^{11}$Be
and $^{12}$Be.
In Fig. \ref{fig:be-rpp}, the ground states are plotted as open circles.
The intruder ground states in $^{11}$Be and $^{12}$Be correspond to 
points on the $\sigma^1$ and $\sigma^2$ lines, respectively.
As shown in the Fig. \ref{fig:be-rpp}, the 2$\alpha$ distances in the ground state of 
$^{11}$Be and $^{12}$Be are about 3 fm.
We may conjecture that the magnitude of 3 fm is the natural distance between
2 $\alpha$, which is energetically favored in these neutron-rich Be isotopes.
Therefore, one of the reasons for the intruder 
ground states can be understood to be a restoration of the natural 
distance at about 3 fm in the very neutron-rich Be isotopes.

The $N=6$ sub-shell closure effects are clearly seen 
in the energy gaps of 6 MeV as well as in the drastic structure changes 
between the $\sigma^0 $ and $\sigma^1$ states in $^{10}$Be.
The sub-shell effect is also reflected in the 
structure change between the $\sigma^1 $ and $\sigma^2$ states in $^{11}$Be,
as shown in the $2\alpha$ distance.
This means that we can regard the vanishing of the magic number 
in neutron-rich Be isotopes
as a shift of the neutron magic number from $N=8$ to $N=6$.

\subsection{Inter-cluster motion and He decay width}

In order to investigate the inter-cluster
motion, we extracted the relative wave functions between clusters 
in the $^{6}$He+$^6$He and $^8$He+$^4$He channels.
We assumed that the intrinsic wave functions of the He clusters are given by
the $0^+$ states of $^4$He, $^{6}$He, $^8$He in the $SU_3$ limit
with the same width parameter as that of $^{12}$Be.
The detailed structures of the $^6$He nucleus, such as the 
neutron halo, are omitted in the present 
analysis for simplicity.
We estimated the inter-cluster motions between clusters in $^{12}$Be by
the reduced width amplitudes ($\tilde y_L(r)$), which were calculated by
projecting the $^{12}$Be wave functions to the cluster model space
expressed by the superpositions of Brink functions.
Moreover, the spectroscopic factors ($\tilde S$) and the 
total cluster probabilities ($\tilde P_c$) are helpful to 
quantitatively discuss the development of He+He clustering.
The detailed definitions and the practical calculation of
$\tilde y_L(r)$, $\tilde S$ and $\tilde P_c$ are explained 
in appendix\ref{app:a}.

\begin{figure}
\noindent
\epsfxsize=0.4\textwidth
\centerline{\epsffile{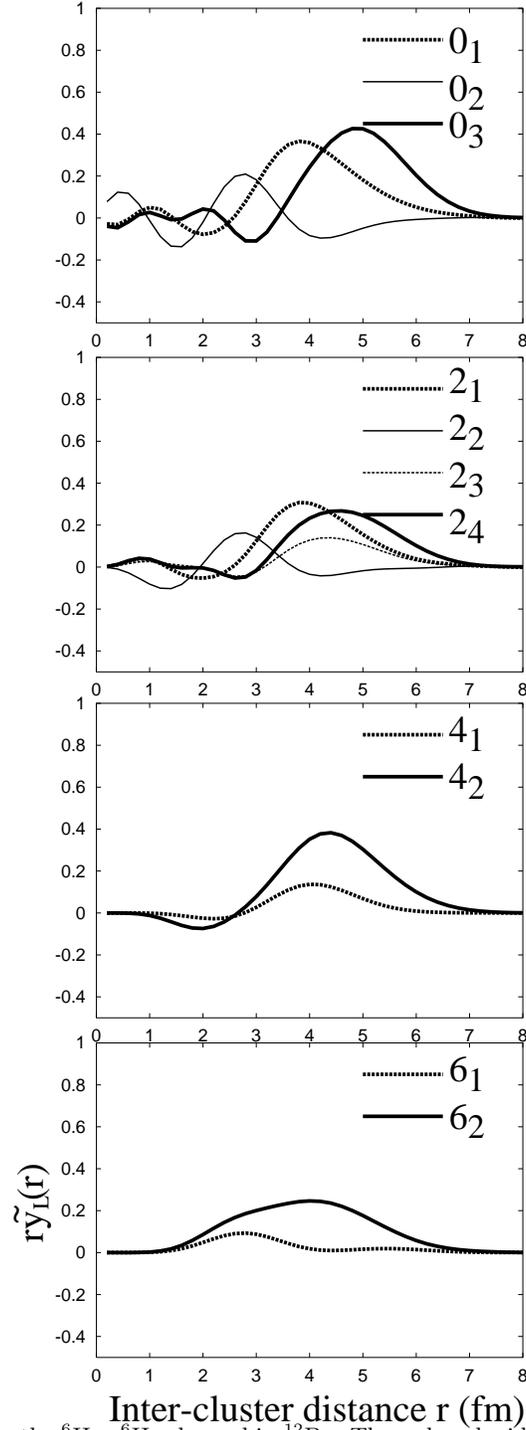}}
\caption{\label{fig:he6}
The inter-cluster motion in the $^6$He+$^6$He channel 
in $^{12}$Be. 
The reduced width amplitudes(R.W.A), $r \tilde y_L(r)$ is defined in the text
and appendix. 
The thick solid(dashed) lines are $r \tilde y_L(r)$ of the states in 
the $K^\pi=0^+_3$($K^\pi=0^+_1$) band.
}
\end{figure}
\begin{figure}
\noindent
\epsfxsize=0.4\textwidth
\centerline{\epsffile{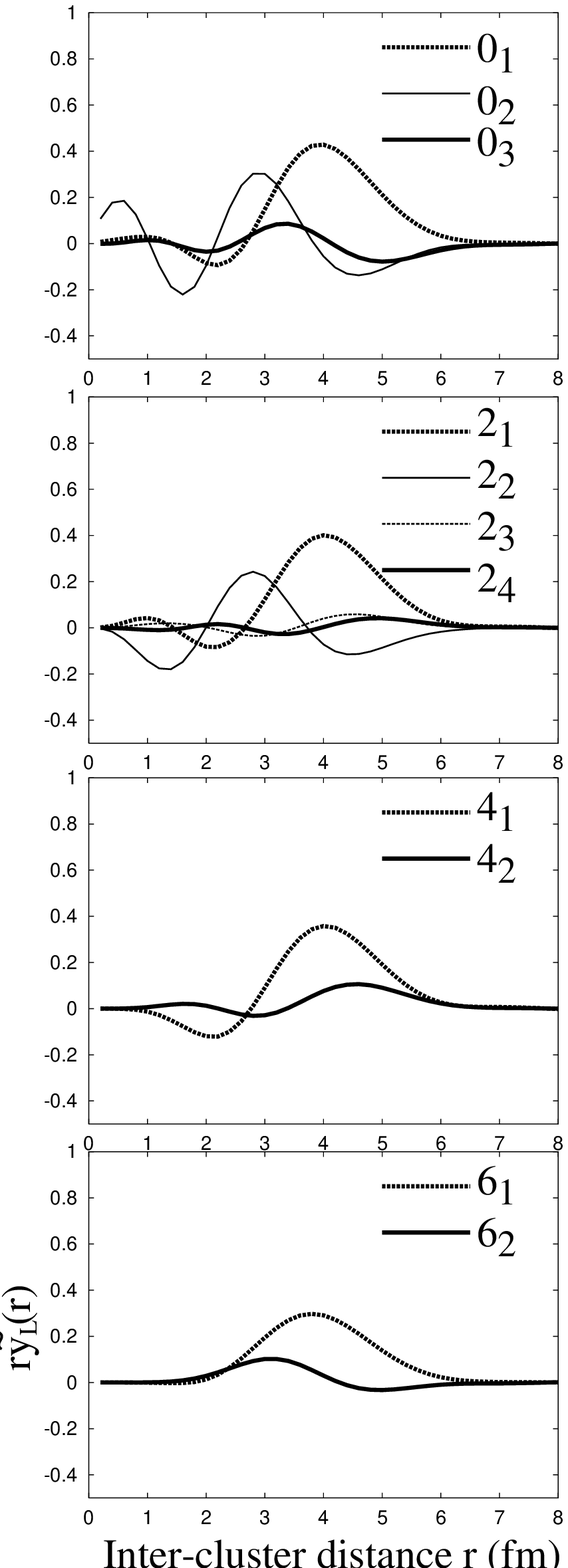}}
\caption{\label{fig:he8}
The inter-cluster motion in the $^8$He+$^4$He channel 
in the ground and excited states of $^{12}$Be. 
The wave function $r \tilde y_L(r)$ is defined in the text and appendix. 
}
\end{figure}

Figs. \ref{fig:he6} and \ref{fig:he8} show the R.W.A. ($r\tilde y_L(r)$)
in the $^6$He+$^6$He and $^8$He+$^4$He channels 
extracted from the obtained $J=L$ states of $^{12}$Be,
respectively. The magnitudes $|r \tilde y_L(r)|$ 
at the channel radius $a=5$ fm, the spectroscopic factors $\tilde S$
and the total cluster probabilities ($\tilde P_c$) 
are presented in Table \ref{tab:be12sfac}.

In each band, even if a low spin state has 
a large amplitude $|r \tilde y_L(r)|$ at the surface region and a 
large spectroscopic factor $\tilde S$, the amplitude becomes smaller
in the high-spin states with $J\ge 6$.
In particular, the amplitudes in the $J=8$ state
are very small. We consider that one of the reason for 
this reduction of the R.W.A. may be the effect of spin-alignments 
in the high spin states.
In the $^6$He+$^6$He channel, 
the amplitude at the surface region is most remarkable 
in the $0^+_3$ state, which has the peak amplitude at the surface
arround $r=5$ fm and a largest spectroscopic factor. 
It indicates the spatially developed $^6$He+$^6$He clustering
in this state. Compared with that in the
$0^+_3$ state, the amplitude in the $0^+_1$ state shifts inward and has
a peak at about $r=4$ fm.
In the $0^+_2$ state, the surface amplitude 
is smallest among three $0^+$ states, which is consistent with this
state has no developed clustering but the dominant neutron $p$-shell closure 
components. The amplitude $|r\tilde y_L(r)|$
at about 4 fm in this $0^+_2$ state are caused by the mixing with the 
$0^+_1$ state. 
Concerning the $J^\pm=2^+_4, 4^+_2,$ and $6^+_2$ states,
both the spectroscopic factors and the surface 
amplitudes $|a y_L(a)| (a=5$ fm) for the $^6$He+$^6$He channel
are not small, but those for the $^8$He+$^4$He channel are small
compared with the $^6$He+$^6$He channel.
On the other hand, with respect to the R.W.A for the $^8$He+$^4$He channel, 
the surface amplitudes at $r=4$ fm are found to be large in the 
states in the $K^\pi=0^+_1$ band, while they are small in other bands such as
$K^\pi=0^+_3$.
By analyzing the spectroscopic factors(Table \ref{tab:be12sfac})
and the R.W.A.(Fig. \ref{fig:he6} and \ref{fig:he8}),
we conclude that the $K^\pi=0^+_1$ band
has both components of $^6$He+$^6$He and $^8$He+$^4$He cluster structures
at least in the low spin states. It is consistent with the result 
by the coupled-channel cluster model with $^6$He+$^6$He and $^8$He+$^4$He
wave functions in Ref. \cite{DESCOUVEMENT}.
On the other hand, in the $K^\pi=0^+_3$ band, the $^6$He+$^6$He 
clustering appears predominantly. 
These results strongly suggested that the neutrons around the 2$\alpha$ core
move about over the whole system in the $K^\pi=0^+_1$ band states, while
in the $K^\pi=0^+_3$ band states they move about not over the whole
system, but around either of the two $\alpha$ clusters.

We next mention the partial decay widths of the excited 
states above the threshold energies of the He decays.
We calculated the theoretical values of the 
partial decay widths ($\Gamma_{^4{\rm He}}$ and 
$\Gamma_{^6{\rm He}}$) concerning the simple binary decays,
$^{6}$He($0^+$)+$^6$He($0^+$) and $^{8}$He($0^+$)+$^4$He($0^+$), using 
the method of reduced width amplitudes:
\begin{eqnarray}
&\Gamma_{^{4,6}{\rm He}}=2P_L(a)\cdot 
\gamma^2_{^{4,6}{\rm He}}(a),\\
&P_L(a)={ka\over F^2_L(ka)+G^2_L(ka)},\\
&\gamma^2_{^{4,6}{\rm He}}(a)={ \hbar^2\over 2\mu a}
|a y_L(a)|^2,  \\
\end{eqnarray}
where $k$ is the wave number of the resonance energy ($E^r$),
$k=\sqrt{2\mu E^r/\hbar^2}$, $a$ is the channel radius, and 
$F_L$ and $G_L$ are the regular and irregular Coulomb functions. 
The resonance energies were evaluated by subtracting
the experimental threshold energies of the channels from the 
theoretical excitation energies. In principle, 
since the present AMD wave functions are not sufficient to describe 
the long tail of the resonance states, we should 
smoothly connect the tail of the irregular Coulomb function ($G_L(kr)$)
to the original relative wave function $r \tilde y_L(r)$ for
the resonance states at the surface region
as usually done in the bound state approximation.  
We found that we can smoothly connect $r \tilde y_L(r)$ with $G_L(kr)$ 
at the point around $r=5$ fm in most of the resonance states 
in the present case. Therefore we chose the channel radius to be $a=5$ fm
to evaluate the partial decay widths.
In Table \ref{tab:be12decay}, the theoretical values of the 
reduced widths $\gamma^2_{{^6}{\rm He}}(a)$, $\gamma^2_{{^4}{\rm He}}(a)$, 
and the partial decay widths are listed. 
The width of the $0^+_3$ state for the $^6$He decay is broad
because of the developed $^6$He+$^6$He cluster structure and the lack of a
centrifugal barrier. 
Although the excited states seem to be stable for the partial
decay width of the He channels, 
the stability of the resonance states should be carefully 
investigated by taking all of the other possible decay channels, 
such as the neutron decays and the excited He decays, into account.

\begin{table}
\caption{ \label{tab:be12sfac} Approximated values for the $S$ factor, 
the total cluster probability of the He-cluster channels in $^{12}$Be.
The magnitudes, $|a \tilde y_L(a)|$, of the reduced width amplitudes
at the channel radius $a=5$ fm are also shown.
}

\begin{center}
\begin{tabular}{c|cc|cc|cc}
     & \multicolumn{2}{c}{$\tilde S$} & \multicolumn{2}{c}{$\tilde P_c$} & 
 \multicolumn{2}{c}{$|a{\tilde y}_L(a)|$} fm$^{-1/2}$ \\ 
 $J^+_n$ & 
{$^6$He+$^6$He} & {$^8$He+$^4$He} & 
{$^6$He+$^6$He} & {$^8$He+$^4$He} & 
{$^6$He+$^6$He} & {$^8$He+$^4$He} \\
\hline
 $0^+_1$ & 0.20 & 0.26 &0.49 & 0.47  &  0.18 & 0.21 \\
 $0^+_2$ & 0.06 & 0.14 & 0.44 & 0.46  &  0.05 & 0.11 \\
 $0^+_3$ & 0.30 & 0.01 &0.37 & 0.02  &  0.43 & 0.08 \\
\hline
 $2^+_1$ & 0.14 & 0.23 & 0.35 & 0.40 &  0.16 & 0.21 \\
 $2^+_2$ & 0.03 & 0.09 & 0.36 & 0.38 &  0.02 & 0.09 \\
 $2^+_3$ & 0.03 & 0.005 &0.05 & 0.01  &  0.11 & 0.05 \\
 $2^+_4$ & 0.13 & 0.003 &0.18 & 0.003  &  0.25 & 0.04 \\
\hline
 $4^+_1$ & 0.02 & 0.19 & 0.04 & 0.28  &  0.07 & 0.19 \\
 $4^+_2$ & 0.24 & 0.02 & 0.38 & 0.02  &  0.30 & 0.09 \\
\hline
 $6^+_1$ & 0.01 & 0.14 & 0.04 & 0.22  &  0.02 & 0.14 \\
 $6^+_2$ & 0.14 & 0.01 & 0.32 & 0.03  &  0.18 & 0.03 \\
\hline
 $8^+_1$ & $\sim 0$ & 0.03 & $\sim 0$ & 0.03  & $\sim 0$ & 0.09 \\
\end{tabular}
\end{center}
\end{table}

\begin{table}
\caption{ \label{tab:be12decay} The theoretical values of the partial 
decay widths and the reduced widths for the $^6$He+$^6$He and 
$^8$He+$^4$He decays
from excited states of $^{12}$Be.
The channel radius is chosen to be $a = 5 $ fm. 
}

\begin{center}
\begin{tabular}{ccc|cc}
     & \multicolumn{2}{c}{$^6$He+$^6$He} &
 \multicolumn{2}{c}{$^8$He+$^4$He} \\ 
 $J^+_n$ & $\gamma^2_{^{6}{\rm He}}(a)$ & $\Gamma_{^{6}{\rm He}}$ (keV) &
$\gamma^2_{^{4}{\rm He}}(a)$ &
$\Gamma_{^{4}{\rm He}}$ (keV) \\
 $0^+_3$ & 2.5 $\times 10^{-1}$ & 7 $\times 10^{2}$ &1.3$\times 10^{-2}$ 
& 4 $\times 10^{1}$\\
 $2^+_4$ & 8.3$\times 10^{-2}$ & 1 & 3.5$\times 10^{-3}$ & 3 \\
 $4^+_2$ & 1.3$\times 10^{-1}$ & 7 & 1.7$\times 10^{-2}$ & 5 \\
 $6^+_2$ & 4.3$\times 10^{-2}$ & 16 & 2.2$\times 10^{-3}$ & 1 \\
 $8^+_1$ & $-$ &    & 1.6$\times 10^{-2}$ & 1 \\
\end{tabular}
\end{center}
\end{table}

\section{Summary}

We studied the structures of the ground and excited states of $^{12}$Be
based on the framework of the AMD method.
This was the first microscopic calculation which systematically reproduced the
energy levels of all of the spin-assigned states in $^{12}$Be,
except for the $1^-$ state. 
One of the present discoveries is the $K^\pi=0^+_3$ band
with the $^6$He+$^6$He cluster structure, 
which corresponds well to the states recently
observed in the He-He break-up reactions.
The ground-state properties, such as the $\beta$-decay strength and
the $E1$ transition strength, were reproduced by the present calculations, 
which suggest that the ground state
is dominated by a deformed intruder state with the developed cluster
structure. The theoretical results predicted many low-lying excited states.
It was found that most of them have a 2$\alpha$ core, while
the $1^+_2$ state has a no $\alpha$-cluster structure.

We analyzed the structures and the single-particle wave functions of 
the intrinsic states, and found rotational 
bands with $K^\pi=0^+_1, 0^+_2, 1^-_1, 0^+_3, 2^+_1$, which are dominated by
the $2\hbar\omega$, $0\hbar\omega$, $1\hbar\omega$, $2\hbar\omega$ and
$2\hbar\omega$ configurations, respectively.  
The $2\hbar\omega$ and $1\hbar\omega$ states have deformed 
structures with a 2$\alpha$ core.
In the positive-parity orbits occupied by the valence neutrons,
we found the molecular orbits, $\sigma$-orbits and the new $\delta'$-orbits. 
The positive-parity molecular orbits play an important role in the 
development of cluster structure. Especially, the reason for the intruder ground state can be understood by the energy gain of the $\sigma$-orbit in the 
developed cluster state.
Although the molecular orbits are helpful to understand the structures of 
the band head states, the model of the 2$\alpha$ and 4 neutrons 
in the molecular orbits is too simple to describe all of the excited states,
because such phenomena as the intrinsic spin alignment, the dissociation 
or breaking of 2$\alpha$ core and the mixing of the molecular orbits 
appear in $^{12}$Be. 

We discussed the inter-cluster motions in the $^6$He+$^6$He and 
$^8$He+$^4$He channels, and estimated the partial width of these decay
channels with the method of reduced width amplitudes.
The analyzed results of the partial widths strongly suggest 
that the neutrons of the
$K^\pi=0^+_3$ band move around either of two $\alpha$ clusters, while the 
neutrons of the $K^\pi=0^+_1$ band move over the whole system.

We also discussed the systematics of the cluster development in the 
neutron-rich Be isotopes according to the classification of the states with
the number of neutrons in the $\sigma$-orbits.
Concerning the vanishing of the neutron magic number $N=8$ 
in the Be isotopes, we suggested that the reason for the 
intruder ground states may be
explained by the restoration of the natural distance between 
the 2 $\alpha$ clusters. 

\acknowledgments

The authors would like to thank Dr. N. Itagaki 
for many discussions. They are also thankful to 
Prof. W. Von Oertzen for helpful comments. 
Valuable comments of Prof. S. Shimoura and Dr. A. Saito
are also acknowledged.
The computational calculations in this work were supported by the 
Supercomputer Project Nos. 58 and 70
of High Energy Accelerator Research Organization(KEK),
and also supported by Research Center for Nuclear Physics
in Osaka University and Yukawa Institute for Theoretical Physics
 in Kyoto University.
This work was partly supported by Japan Society for the Promotion of 
Science and a Grant-in-Aid for Scientific Research of the Japan
Ministry of Education, Science and Culture.
This work was partially performed in the ``Research Project for Study of
Unstable Nuclei from Nuclear Cluster Aspects'' sponsored by
Institute of Physical and Chemical Research (RIKEN).

\appendix{
\section{Inter cluster wave functions}\label{app:a}

The calculational methods of  
the reduced width amplitudes(R.W.A.), spectroscopic factors, and 
total cluster probability concerning the $^6$He(0$^+$)+$^6$He(0$^+$) and
$^8$He(0$^+$)+$^4$He(0$^+$) channels are described in this appendix.

\subsection{Reduced width amplitudes and cluster model space}
The reduced width amplitudes, $y_L(a)$, are defined as follows:
\begin{eqnarray}
&y_L(a)\equiv q_1
\langle \frac{\delta(r-a)}{r^2} Y_{L0}(\hat r)\phi_0({\rm C}_1)
\phi_0({\rm C}_2)
| \Phi_L \rangle,\\
&q_1\equiv 
{1\over \sqrt{1+\delta_{C_1,C_2}}} 
\sqrt{\left(\begin{array}{c} A \cr A_1 \end{array}\right)},
\end{eqnarray}
where $\Phi_L$ is the internal wave function of a model wave function
and $\phi_0({\rm C}_1)$ and $\phi_0({\rm C}_2)$ 
are the internal wave functions of 
the clusters C$_1$ and C$_2$. 
The mass numbers of the system and the clusters ${\rm C}_1$ 
and ${\rm C}_2$
are $A$, $A_1$ and $A_2$, respectively.

Here, we briefly review the R.W.A. in the system 
of two clusters written in the form of 
RGM(resonating group method)
or GCM(generator coordinate method) wave functions.
The detailed calculational methods of the RGM and GCM kernels are 
described, for example, in Ref.\cite{HORIUCHIgcm}. 
We assume the same width parameter ($\nu$) for clusters
${\rm C}_1$ and ${\rm C}_2$ described by harmonic oscillator shell model 
wave functions.
When $\Phi_{L}$ is an RGM-type wave function expressed as,
\begin{eqnarray}
&\Phi_{L}=q_2
{\cal A}\{\chi_L(r) Y_{L0}(\hat r)\phi_0({\rm C}_1)\phi_0({\rm C}_2)\},
\label{eqn:rgm}\\
&q_2\equiv \frac{1}{\sqrt{(1+\delta_{{\rm C}_1,{\rm C}_2})
\left(\begin{array}{c} A \\ A_1 \end{array}\right)}},
\end{eqnarray}
the R.W.A., $y_L(a)$, can be calculated based on the knowledge of 
the RGM norm kernel as follows.
By expanding the relative motion $\chi_L(r)$ with the radial 
harmonic oscillator (H.O.) wave functions $R_{nl}(r,\nu')$
with the width 
parameter $\nu'=\frac{A_1 A_2}{A}\nu$ as
\begin{equation}\label{eqn:chi}
\chi_L(r)=\sum_n e_{nL} R_{nL}(r,\nu'),
\end{equation}
we obtain the R.W.A. as,
\begin{equation}\label{eqn:yl}
y_L(a)=\sum_n e_{nL} \mu_{nL} R_{nL}(a,\nu'),
\end{equation}
where $\mu_{nL}$ are the eigen values of the RGM norm kernel.
If the RGM-type wave function \ref{eqn:rgm} is normalized to unity, 
the following equation is satisfied:
\begin{equation}\label{eqn:unit}
\sum_n e^2_{nL} \mu_{nL}=1.
\end{equation}

In the case of general A-nucleon wave functions, $\Phi_L$ can be separated 
into the cluster part and the non-cluster part,
\begin{equation}\label{eqn:projcl}
\Phi_L={\cal A}\{\chi_L(r) Y_{L0}(\hat r)\phi_0({\rm C}_1)\phi_0({\rm C}_2)\}+\Phi^R_L,
\end{equation}
where $\Phi^R_L$ is the residual part after projection of the 
cluster model space, and satisfies 
$\langle Y_{L0}(\hat r)\phi_0({\rm C}_1)\phi_0({\rm C}_2)|\Phi^R_L\rangle=0$.
In this case, the R.W.A. of $\Phi_L$ can be
obtained with equation \ref{eqn:yl} by using an expansion of 
$\chi_L(r)$ with the H.O. functions shown in Eq.\ref{eqn:chi}.
The normal spectroscopic factors $S$ are calculated as
\begin{equation}\label{eqn:spec}
S=\int y_L(r)^2 r^2 dr=\sum_n \mu^2_{nL} e^2_{nL}.
\end{equation}
It is useful to consider the total probability of the clustering 
component (total cluster probability) 
in $\Phi_L$ defined by the squared overlap with the 
cluster model space calculated as,
\begin{equation}\label{eqn:pc}
P_c=\sum_n \mu_{nL} e^2_{nL}.
\end{equation}
It is easily found that $P_c=1$ for the normalized RGM-type wave functions 
 from equation \ref{eqn:unit}.

The R.W.A. of the wave function
$\Phi_L(^{12}{\rm Be})$ in the AMD framework is 
\begin{equation}
y_L(a)\equiv 
q_1\langle \frac{\delta(r-a)}{r^2} Y_{L0}(\hat r)\phi_0({\rm C}_1)
\phi_0({\rm C}_2)
| \Phi_L(^{12}{\rm Be}) \rangle,
\end{equation}
where $\Phi_L(^{12}{\rm Be})$ is a normalized spin-parity projected 
AMD wave function, where the center of mass motion 
is extracted,
and $\phi_0({\rm C}_n)$ (${\rm C}_n=^4$He, $^6$He, $^8$He)
are the intrinsic wave functions of the $SU_3$-limit $0^+$ states of 
the He clusters. 
In order to project to the cluster model space from the $^{12}$Be
wave functions ($\Phi_L(^{12}{\rm Be})$) and calculate the radial 
wave functions
 ($\chi_L(r)$) between the clusters, as mentioned in Eq.\ref{eqn:projcl},
we use the orthonormal sets of the GCM wave functions.
Concerning the $k$-th cluster wave functions $\Psi^c({\bf b}_k)$ with 
the inter-cluster distances \{$b_k$\},
we adopt Brink-type functions, where two clusters(${\rm C}_1=^{A_1}$He and 
${\rm C}_2=^{A_2}$He) written in the $SU_3$-limit representation 
are located at the points $(0, 0, \frac{A_2}{A_1+A_2}b_k)$ and
$(0, 0, -\frac{A_1}{A_1+A_2}b_k)$, respectively,
\begin{equation}
\Psi^c({\bf b}_k)= {\cal A} \{
\Psi({\rm C}_1,\frac{A_2}{A}{\bf b}_k)
\Psi({\rm C}_2,-\frac{A_1}{A}{\bf b}_k)
\}, \qquad {\bf b}_k=(0,0,b_k).
\end{equation}
The width parameters of the He clusters are chosen to be the
same as that of the $^{12}$Be wave functions for simplicity.
Then we can rewrite $\Psi^c({\bf b}_k)$ as
\begin{eqnarray}
&\Psi^{c}({\bf b}_k)=\omega_0({\bf X}_G)\cdot \Phi^{c}({\bf b}_k),\\
&\Phi^{c}({\bf b}_k)=
q_2{\cal A}\lbrace \Gamma({\bf r},{\bf b}_k, \nu') \phi_0({\rm C}_1)
\phi_0({\rm C}_2)\rbrace,
\\
& \Gamma({\bf r},{\bf b}_k, \nu')=(\frac{2\nu'}{\pi})^{3/4}
e^{-\nu'({\bf r}-{\bf b}_k)^2}\\
&\omega_0({\bf X}_G)\equiv \left( \frac{2A\nu}{\pi} \right)^{3/4}
e^{-A\nu {\bf X}_G^2}, \quad{\bf X}_G\equiv \frac{1}{A}\sum_{i=1,A} {\bf r}_i,
\end{eqnarray}

In each $L$, we make a set of the 
orthonormal basis $\tilde \Phi_{k,L}$ by linear combinations of 
the total-spin-projected GCM wave functions $\Phi_{k,L}$:
\begin{eqnarray}
&\Phi_{k,L}\equiv q_{kL}
{P^L_{00}\Phi^{c}({\bf b}_k)},\\
&\tilde \Phi_{k,L}=\sum_{k'}A^{(L)}_{kk'}\Phi_{k',L},\\
&\langle \tilde \Phi_{k,L}|\tilde \Phi_{k',L} \rangle =\delta_{kk'},\\
\end{eqnarray}
where $q_{kL}$ are the normalization factors $q_{kL}\equiv 1/
{\langle P^L_{00}\Phi^{c}({\bf b}_k)| P^L_{00}\Phi^{c}({\bf b}_k)
\rangle ^{1/2} }.$
By using the 
partial wave expansion of the functions 
$\Gamma({\bf r},{\bf b}_k,\nu')$, the radial functions $\chi^{(k)}_L(r)$
in the $k$-th wave functions $\Phi_{k,L}$ can be expressed as follows:
\begin{eqnarray}
& \Phi_{k,L}
=\frac{1}{\sqrt{(1+\delta_{{\rm C}_1,{\rm C}_2})
\left(\begin{array}{c} A \\ A_1 \end{array}\right)}}
{\cal A}\lbrace \chi^{(k)}_L(r) Y_{L0}(\hat r)
 \phi_0({\rm C}_1)\phi_0({\rm C}_2)\rbrace,\\
& \chi^{(k)}_L(r)=q_{kL}
\sqrt{\frac{2l+1}{4\pi}}
\Gamma_L(r,b_k,\nu'),\\
&\Gamma_L(r,b_k,\nu') =
\left(\frac{2\nu'}{\pi}\right)^{3/4}4\pi i_L(2\nu' r b_k)e^{-\nu'(r^2+b_k^2)},
\end{eqnarray}
where $i_L$ is the modified spherical Bessel function.
We assume that the projection operator $P^c_L$ onto the cluster 
model space can 
be written by the orthonormal basis $\tilde \Phi_{k,L}$,
\begin{equation}
P^c_L= 
\sum_k |\tilde \Phi_{k,L}\rangle\langle \tilde \Phi_{k,L}|
=\sum_{k,k',k''} |\Phi_{k',L}\rangle A^{(L)}_{kk'}A^{(L)*}_{kk''}\langle
\Phi_{k'',L}|.
\end{equation}
In this case, the radial function $\chi_L(r)$ in Eq.\ref{eqn:projcl}
for $\Phi_L(^{12}{\rm Be})$ is written as
\begin{equation}\label{eqn:chibe12}
\chi_L(r)= \sum_{k'} \chi^{(k')}_L(r)
\{\sum_{k,k''}  A^{(L)}_{kk'}A^{(L)*}_{kk''}\langle
\Phi_{k'',L}|\Phi_L(^{12}{\rm Be})\rangle\}.
\end{equation}

\subsection{Practical calculation of $\mu_{nL}$}

In the present calculations, 
we chose $b_k=1,2,\cdots,9$ fm for $L=0,2$ ($L=0,2,4,6$), 
$b_k=2,3\cdots,9$ fm for $L=4,6$,  
and $b_k=3,\cdots,9$ fm for $L=8$ ($L=8$)
for the $^8$He+$^4$He ($^6$He+$^6$He) channel.
In the practical calculations, we approximate $\mu_{nL}$ with  
$\tilde \mu_{nL}$ calculated as follows.
It is known that the eigen values $\mu_{nL}$ equal to zero for 
the Pauli forbidden states with $N=2n+L < N_{min}$ 
($N_{min}$ is the minimum allowed number)
and $\mu_{nL}\sim 1$ with enough large numbers, $N=2n+L$.
In order to obtain approximated values $\tilde \mu_{nL}$,
we truncate the quanta $n$ 
with a finite number $n=n_{min},\cdots,n_{max} 
(N_{min}=2n_{min}+L,$ $N_{max}=2n_{max}+L)$ 
by assuming $\tilde \mu_{nL}=1$ for $N=2n+L > N_{max}$.
With the use of the expansion of $\chi^{(k)}_L(r)$ in the normalized 
cluster wave functions $\Phi_{k,L}$ by $R_{nL}(r,\nu')$
\begin{equation}
\chi^{(k)}_L(r)=\sum_n e^{(k)}_{nL} R_{nL}(r,\nu'),
\end{equation}
we obtain $\tilde \mu_{nL}$ by following equations:
\begin{equation}\label{eqn:appmu}
\sum_n e^{(k)2}_{nL}\tilde \mu_{nL}=1.
\end{equation}
The approximated values 
$\tilde \mu_{nL}$ for simple systems, $\alpha+\alpha$ and 
$\alpha$+$^{16}$O are compared with 
the exact eigen values $\mu_{nL}$ in Table \ref{tab:mutilde}.

\begin{table}
\caption{ \label{tab:mutilde} The eigen values $\mu_{N=2n+L}$ 
and approximated 
values ${\tilde \mu_{nL}} (L=0)$ of the RGM norm 
kernel for $\alpha+\alpha$ and $\alpha$+$^{16}$O systems.
The values $\mu_{N}$ are taken from Ref.\protect\cite{HORIUCHIgcm}.
The width parameters for $\alpha$ and $^{16}$O are assumed to be same.}
\begin{center}
\begin{tabular}{c|c|c||c|c|c}
\multicolumn{3}{c}{$\alpha+\alpha$}&
\multicolumn{3}{c}{$\alpha+^{16}$O}\\
$N$& $\mu_{N}$ & ${\tilde \mu_{nL}}$ &$N$& $\mu_{N}$ & $\tilde \mu_{nL}$\\ 
\hline
0&0&0&0&0&0\\
2&0&0&2&0&0\\
4&0.7500 &0.7500 &4&0&0\\
6&0.9375 &0.9375 &6&0&0\\
8&0.9844 &0.9844 &8&0.2292&0.2292\\ 
10&0.9961&0.9962 &10&0.5103&0.5063 \\
12&0.9990&0.9987 &12&0.7185&0.7538 \\
&&&14&0.8459&0.7399 \\
&&&16&0.9178&1\\
&&&18&0.9568&1\\
&&&20&0.9775&1\\
&&&22&0.9884&1\\
&&&24&0.9941&1\\
&&&26&0.9970&1\\
\end{tabular}
\end{center}
\end{table}

In the present calculation of $^{12}$Be, $N_{max}$ was chosen to be 14.
By using $\tilde \mu_{nL}$ determined by Eq.\ref{eqn:appmu},
we define the approximate R.W.A., $\tilde y_L(a)$, 
the spectroscopic facors, $\tilde S$, and the total cluster probability, 
$\tilde P_c$ for $\Phi_L(^{12}{\rm Be})$ 
by analogy with the relations in Eqs. \ref{eqn:yl}, \ref{eqn:spec} 
and \ref{eqn:pc}:
\begin{eqnarray}
&\tilde y_L(a)\equiv \sum_n e_{nL} \tilde \mu_{nL} R_{nL}(a,\nu'),\\
&\tilde S\equiv \sum_n \tilde \mu^2_{nL} e^2_{nL},\\
&\tilde P_c \equiv \sum_n \tilde \mu_{nL} e^2_{nL},
\end{eqnarray}
where the coefficients $e_{nL}$ were calculated by
$e_{nL}=\int r^2 \chi_L(r)R_{nL} dr$ from $\chi_L(r)$ given in 
Eq. \ref{eqn:chibe12}. 
}

\section*{References}
\begin{thebibliography}{9}
  
\bibitem{SEYA}
 M. Seya, M. Kohno, and S. Nagata, Prog. Theor. Phys.
 {\bf 65}, 204 (1981).
\bibitem{ENYObc}
 Y. Kanada-En'yo, H. Horiuchi and A. Ono,
Phys. Rev. C {\bf 52}, 628 (1995);
 Y. Kanada-En'yo and H. Horiuchi,
Phys. Rev. C {\bf 52}, 647 (1995).
\bibitem{OERTZEN}
W. von Oertzen, Z. Phys. A {\bf 354}, 37 (1996); {\bf 357}, 355(1997).
\bibitem{ARAI}
K. Arai, Y. Ogawa, Y. Suzuki and
 K. Varga , Phys. Rev. C {\bf 54}, 132 (1996). 
\bibitem{DOTE}
A. Dot\'{e}, H. Horiuchi, and Y. Kanada-En'yo, 
Phys. Rev. C {\bf 56}, 1844 (1997).
\bibitem{ENYOf}
 Y. Kanada-En'yo, H. Horiuchi and A. Dot\'{e},
J. Phys. G, Nucl. Part. Phys. {\bf 24} 1499 (1998).
\bibitem{ENYOg}
Y. Kanada-En'yo, H. Horiuchi and A. Dot\'{e}, Phys. Rev. {\bf C 60}, 
064304(1999).
\bibitem{ITAGAKI}
N. Itagaki and S. Okabe, Phys. Rev. C {\bf 61}, 044306 (2000);
\bibitem{ITAGAKIa}
N. Itagaki, S. Okabe and K. Ikeda, Phys. Rev. C {\bf 62}, 034301 (2000).
\bibitem{OGAWA}
Y. Ogawa, K. Arai, Y. Suzuki, and K. Varga, Nucl. Phys. {\bf A673}
122 (2000).
\bibitem{ENYObe11}
Y. Kanada-En'yo and H. Horiuchi, Phys. Rev. C {\bf 66}, 024305(2002).
\bibitem{ENYObe14}
Y. Kanada-En'yo, Phys. Rev. {\bf C 66}, 011303(2002).
\bibitem{TANIHATA}
 A.A. Korsheninnikov, et al., Phys. Lett. {\bf B} 343, 53(1995).
\bibitem{FREER}
M. Freer, et al., Phys. Rev. Lett. {\bf 82}, 1383 (1999);
M. Freer, et al., Phys. Rev. C {\bf 63}, 034301 (2001).
\bibitem{SAITO}
A. Saito, et al., {\it Proc. Int. Sympo. on 
        Clustering Aspects of Quantum Many-Body Systems}, eds
A. Ohnishi, N. Itagaki, Y. Kanada-En'yo and K. Kato,
(World Scientific Publishing Co.).
\bibitem{ITO}
M. Ito, Y. Sakuragi, Y. Hirabayashi, Phys.Rev.{\bf C 63}, 064303(2001). 
\bibitem{DESCOUVEMENT}
P. Descouvement and D. Baye, Phys. Lett. B {\bf 505}, 71(2001).
\bibitem{BARKER}
F. C. Barker, J. Phys. G, Nucl. Part. Phys. {\bf 2}, L45 (1976).
\bibitem{FORTUNE}
H. T. Fortune and G.-B. Liu, Phys. Rev. {\bf C} 50, 1355 (1994).
\bibitem{TSUZUKIa}
T. Suzuki and T. Otsuka, Phys. Rev. {\bf C} 56, 847(1997).
\bibitem{IWASAKI}
H. Iwasaki, et al. Phys.Lett.{\bf B}491, 8(2000).
\bibitem{SHIMOURA}
S. Shimoura, et al., preprint, CNS-REP-47(2002).
\bibitem{NUNES}
F. M. Nunes, I. J. Thompson, J. A. Tostevin, Nucl. Phys. {\bf A703},
 593 (2002). 
\bibitem{OKABE}
S. Okabe, Y. Abe, and H. Tanaka, Prog. Theor. Phys.{\bf 57}, 866(1977);
S. Okabe, Y. Abe, Prog. Theor. Phys.{\bf 61}, 1049(1979).
\bibitem{ENYOa}
 Y. Kanada-En'yo and H. Horiuchi, Prog. Theor. Phys.
 {\bf 93}, 115 (1995).
\bibitem{ENYOe}
 Y. Kanada-En'yo,
Phys. Rev. Lett. {\bf 81}, 5291 (1998).
\bibitem{TOHSAKI}
 T. Ando, K.Ikeda, and A. Tohsaki, Prog. Theor. Phys.
 {\bf 64}, 1608 (1980).
\bibitem{LS}
 N. Yamaguchi, T. Kasahara, S. Nagata, and Y. Akaishi,
 Prog. Theor. Phys. {\bf 62}, 1018 (1979);
 R. Tamagaki, Prog. Theor. Phys. {\bf 39}, 91 (1968).
\bibitem{TAN88}
I. Tanihata et al., Phys.Lett.B 206, 592(1988).
\bibitem{CHOU}
W.-T. Chou, E. K. Warburton and B. A. Brown, Phys. Rev. C{\bf 47},
163(1993).
\bibitem{OERTZENa}
W. von Oertzen, Nuovo Cimento 110, {\bf 895}, (1997).
\bibitem{ENYOsup}
Y. Kanada-En'yo and  H. Horiuchi, Prog. Theor. Phys. Suppl.{\bf 142},
 205(2001).
\bibitem{HORIUCHIgcm}
H. Horiuchi, Chapter III of Suppl. of Prog. Theor. Phys. No.62(1977). 
\end{thebibliography}
\end{document}